\documentclass[aip,sd,amsmath,amssymb,preprint]{revtex4-1}
\usepackage{graphicx}
\usepackage{dcolumn}
\usepackage{bm}
\DeclareMathOperator\arctanh{arctanh}
\usepackage{comment}
\usepackage[usenames,dvipsnames]{xcolor}
\newcommand{\oldversion}[1]{}

\begin{document}
\title{Differential Capacitance of Ionic Liquids According to Lattice-Gas Mean-Field Model with Nearest-Neighbor Interactions}

\author{Rachel Downing}
\affiliation{Department of Physics, North Dakota State University, Fargo ND 58108, USA}
 
\author{Bjorn K.~Berntson}
\affiliation{Department of Mathematics, North Dakota State University, Fargo ND 58108, USA}

\author{Guilherme V.~Bossa}
\affiliation{Department of Physics, S\~ao Paulo State University (UNESP), Institute of Biosciences, Humanities and Exact Sciences, S\~ao Jos\'e do Rio Preto, SP,15054-000, Brazil}

\author{Sylvio May} 
\affiliation{Department of Physics, North Dakota
  State University, Fargo ND  58108, USA}

\date{\today}

\begin{abstract}
The Bragg-Williams free energy is used to incorporate nearest-neighbor interactions into the lattice gas model of a solvent-free ionic liquid near a planar electrode. We calculate the differential capacitance from solutions of the mean-field consistency relation, arriving at an explicit expression in the limit of a weakly charged electrode. The two additional material parameters that appear in the theory -- the degree of nonideality and the resistance to concentration changes of each ion type -- give rise to different regimes that we identify and discuss. As the nonideality parameter, which becomes more positive for stronger nearest-neighbor attraction between like-charged ions, increases and the electrode is weakly charged, the differential capacitance is predicted to transition through a divergence and subsequently adopt negative values just before the ionic liquid becomes structurally unstable. This is associated with the spontaneous charging of an electrode at vanishing potential. The physical origin of the divergence and the negative sign of the differential capacitance is a nonmonotonic relationship between surface potential and surface charge density, which reflects the formation of layered domains alternatingly enriched in counterions and coions near the electrode. The decay length of this layered domain pattern, which can be many times larger than the ion size, is reminiscent of the recently introduced concept of ``underscreening''. 
\end{abstract}

\maketitle

\section{Introduction}
Ionic liquids consist of molten salts \cite{wishart09,armand11} that stay in their liquid phase at ambient temperatures \cite{watanabe17}. They avoid transitioning into a crystal phase by being of sufficient molecular size to reduce Coulomb interactions and by displaying packing incompatibilities that disfavor a stable crystal structure \cite{hayes15,wishart09}. Ionic liquids tend to exhibit high thermal stability \cite{watanabe17} and resistance to both combustion and evaporation due to their low vapor pressure \cite{wishart09}. Among their many applications \cite{balducci07,armand09,macfarlane14,watanabe17,watanabe17,salanne17} are batteries, fuel cells, and supercapacitors.

In the vicinity of a charged electrode, ionic liquids form an electric double layer (EDL) that reflects the competition between electrostatic attraction of the counterions to the electrode surface and the translational entropy of the ions. However, in contrast to electrolytes that are diluted by solvent molecules, the ions of an ionic liquid exhibit strongly nonideal mixing properties due to electrostatic correlations and non-electrostatic ion-ion interactions, and can therefore not be described on the basis of the classical Poisson-Boltzmann theory. The pronounced nonideality has implications for the structure of the EDL, which can be characterized and discussed on the basis of the differential capacitance \cite{fedorov14}.

One of the most commonly used mean-field models to describe the structure of the EDL formed by an ionic liquid is based on a lattice gas approximation for the translational entropy of the ions \cite{bikerman42,kornyshev07}, including extensions to asymmetric ion sizes \cite{popovic13,han14} and comparisons with Monte Carlo simulations \cite{girotto17}. Its simple prediction for the differential capacitance often serves as a reference for the interpretation of experimental and computational results \cite{lockett10,lamperski09,caetano16}. Equations of state other than that for the lattice gas have been studied \cite{lue99,biesheuvel07}, including formalisms that are valid for a general form of the underlying equation of state \cite{gavish16,maggs16}. To account for short-ranged correlations between the ions of an ionic liquid, Bazant {\em et al} \cite{bazant11} have proposed a phenomenological extension of the mean-field free energy of an ionic liquid. This extension, which gives rise to ion crowding and overscreening, has triggered a multitude of follow-up investigations of how short-ranged interactions and correlations influence the structure of the EDL \cite{aoki12,caetano16,goodwin17,gavish17,bokun18,yin18,rotenberg18}. They are also inspired by the experimental finding of a large screening length in concentrated electrolytes \cite{perkin12,gebbie13,gebbie17,perez17}. 

Some of the recently proposed phenomenological modeling approaches incorporate short-range ion-ion interactions on the basis of adding coupling terms between anion and cation concentrations \cite{goodwin17,kornyshev17} plus additional gradient terms \cite{blossey17,gavish17} to the underlying free energy. As in Landau theory, gradient terms are generally present in inhomogeneous systems such as an EDL \cite{blossey17}. While certain aspects -- like the influence of the coupling terms on the differential capacitance in the absence of gradient terms \cite{goodwin17} or the emergence of a damped oscillating potential with a large screening length in the presence of both coupling and gradient terms \cite{gavish17} -- have been discussed in previous work, there is no comprehensive analysis yet of how coupling and gradient terms jointly affect the differential capacitance of an ionic liquid and how this relates to oscillations and the screening length of the electrostatic potential. To provide such a detailed analysis is the goal of the present work.

Here, we incorporate short-range ion-ion interactions into the lattice gas description of an incompressible ionic liquid on the level of mean-field theory employing the Bragg-Williams free energy \cite{davis96}, thereby including both coupling and gradient terms. We point out that this model still operates on the mean-field level because nearest-neighbor interactions are described using a random mixing approximation without accounting for correlations \cite{doi13}. Approaches for general underlying equations of state have been proposed recently \cite{gavish16,maggs16}, but we focus exclusively on the lattice gas because it leads to a simple analytic expression for the differential capacitance in the limit of a weakly charged electrode.

The main objective of our work is to provide a quantitative analysis and discussion of how and when additional coupling and gradient terms in the underlying lattice gas free energy give rise to oscillating potentials with large screening lengths and how this is reflected in the differential capacitance. Specifically, we shall demonstrate that at highly nonideal conditions (that is, large coupling terms), where the ionic liquid approaches a structural instability, the differential capacitance is predicted to transition through a divergence and then to become negative or to induce a spontaneous charging of the electrode \cite{partenskii08,partenskii11}. Divergence and sign reversal arise from a nonmonotonic behavior of the surface potential as function of the electrode's surface charge density, accompanied by a spatially oscillating potential that causes the formation of layered domains alternatingly enriched in counterions and coions near the electrode, also known as ``overscreening'' \cite{bazant11}. The layered domain pattern can propagate far into the ionic liquid, a property associated with the term ``underscreening'' \cite{perez17,gebbie17}.

\section{Theory}
We consider a lattice gas model to describe the structure of the EDL that is formed by an incompressible, solvent-free ionic liquid, composed of monovalent ions, in the vicinity of a planar electrode of fixed surface charge density $\sigma$. The lateral area $A$ of the electrode is sufficiently large so that edge effects can be neglected and, on the mean field level, all physical quantities become functions of the distance $x$ to the electrode only. Anions and cations are of the same size, each occupying one single lattice site of volume $\nu$. The absence of solvent requires that the local volume fractions of anions (index ``$a$'') and cations (index ``$c$''), $\phi_a=\phi_a(x)$ and $\phi_c=\phi_c(x)$, add up to $\phi_a+\phi_c=1$ at each position $x$. In the bulk of the ionic liquid, at $x \rightarrow \infty$, fluidity and charge neutrality require $\phi_a(x \rightarrow \infty)=\phi_c(x \rightarrow \infty)=1/2$. We express the mean-field free energy per unit area of the ionic liquid as $F/A=(1/\nu) \int_0^\infty dx f$, with the local free energy per lattice site
\begin{equation} \label{sw31}
f=\frac{l^2}{2} \Psi'^2+\phi_a \ln \phi_a+\phi_c \ln \phi_c+\chi \phi_a \phi_c+\frac{\alpha}{2} l^2 (\phi_a'^2+\phi_c'^2).
\end{equation}
Here, $\Psi=\Psi(x)$ denotes the scaled electrostatic potential, following the common definition $\Psi=e \Phi/k_BT$, with the electrostatic potential $\Phi=\Phi(x)$, elementary charge $e$, Boltzmann constant $k_B$, and absolute temperature $T$. A prime denotes the derivative, i.e.~$\Psi'=d\Psi/dx$, $\Psi''=d^2\Psi/dx^2$, $\phi_a'=d\phi_a/dx$, and $\phi_c'=d\phi_c/dx$. Electrostatic interactions are characterized by the length scale $l=\sqrt{\nu/(4 \pi l_B)}$, where $l_B=e^2/(4 \pi \epsilon \epsilon_0 k_BT)$ is the Bjerrum length that reflects the dielectric constant $\epsilon$ of the ionic liquid ($\epsilon_0$ is the permittivity of free space). Hence, the term $l^2 \Psi'^2/2$ in Eq.~\ref{sw31} accounts for the energy, per lattice site, stored in the electric field. Note that the length $l$ tends to be much smaller than the size of an ion. For example, $\nu$ $=1 \: \mbox{nm}^3$ and $l_B=7 \: \mbox{nm}$ yield $l=0.1 \: \mbox{nm}$.

All terms following the first one on the right-hand side of Eq.~\ref{sw31} constitute the Bragg-Williams free energy density of an inhomogeneous lattice gas \cite{davis96}, which accounts for nearest-neighbor interactions. The two terms $\phi_a \ln \phi_a$ and $\phi_c \ln \phi_c$ in Eq.~\ref{sw31} express the mixing entropy contribution of a lattice gas, consisting of anions and cations. The fourth term in Eq.~\ref{sw31}, $\chi \phi_a \phi_c$, is a coupling term that characterizes the extent of nearest-neighbor interactions between the ions according to the random mixing approximation \cite{davis96}. The degree of nonideality $\chi=z [\omega_{ac}-(\omega_{aa}+\omega_{cc})/2]$ reflects the anion-cation ($\omega_{ac}$), anion-anion ($\omega_{aa}$), and cation-cation ($\omega_{cc}$) interaction strengths, and $z$ is the lattice coordination number. Both non-electrostatic and electrostatic interactions may enter into $\chi$, the latter for example through correlations or dipole (or higher) electrostatic moments that are not accounted for by the term $l^2 \Psi'^2/2$ in Eq.~\ref{sw31}. Note that $\chi>0$ implies an effective repulsion between anions and cations, and $\chi>2$ may lead to a macroscopic phase transition in an (hypothetically) uncharged bulk system. The last term in Eq.~\ref{sw31}, $\alpha l^2 (\phi_a'^2+\phi_c'^2)/2$, is a gradient term that describes the energy penalty due to compositional changes, where $\alpha$ (with $\alpha \ge 0$) is a dimensionless constant. Note that a more general approach would assign two different prefactors, $\alpha_a$ and $\alpha_c$, to the terms $\alpha_a l^2 \phi_a'^2$ and $\alpha_c l^2 \phi_c'^2$. The relations of these two prefactors to the molecular interaction strengths are $\alpha_a= (\nu^{1/3}/l)^2 (\omega_{ac}-\omega_{aa})$ and $\alpha_c= (\nu^{1/3}/l)^2 (\omega_{ac}-\omega_{cc})$. Hence, our present work assumes $\omega_{aa}=\omega_{cc}$, which implies $\alpha=\alpha_a=\alpha_c$. In Eq.~\ref{sw31} we neglect the presence of terms that contain higher order derivatives such as $\phi_a''^2$ and $\phi_c''^2$. Note that Eq.~\ref{sw31} properly accounts for the equilibrium with a bulk ionic liquid (at $x \rightarrow \infty$) where $\phi_a=\phi_c=1/2$. Also, in Eq.~\ref{sw31} and in the remainder of this paper, we express energies in units of the thermal energy $k_BT$.

The free energy in Eq.~\ref{sw31} represents the incorporation of short-range ion-ion interactions on the mean-field level within the framework of the lattice gas model. It is the sole basis of the present work, and we provide an in-depth analysis of its implications for the electrostatic potential, structure of the EDL, and differential capacitance. Eq.~\ref{sw31} is similar to lattice gas approaches with additional nearest-neighbor interactions used in previous studies. Goodwin {\em et al} \cite{goodwin17} as well as Yin {\em et al} \cite{yin18} have accounted for the three distinct molecular interaction parameters $\omega_{aa}$, $\omega_{cc}$, and $\omega_{ac}$ explicitly for a compressible lattice gas, yet without including gradient terms. Goodwin and Kornyshev \cite{kornyshev17} have proposed a similar model with the addition of another species of ``spectating'' ions. Gavish {\em et al} \cite{gavish17} use a free energy of an electrolyte in the presence of solvent with the same interaction model as in our present work. Their work focuses on the screening length as function of dilution but not on the differential capacitance. Finally, Blossey {\em et al} \cite{blossey17} use a general form of the underlying equation of state, including all compositional gradient terms. Upon linearizing the self-consistency relation they demonstrate the ability of the potential to undergo damped oscillations, yet without analyzing the differential capacitance. 

We minimize the free energy $F$ and find the equilibrium distributions for $\phi_a$ and $\phi_c$. To this end, we note the Poisson equation $l^2 \Psi''=\phi_a-\phi_c$ that relates the second derivative of the potential $\Psi$ to the difference of the local anion and cation volume fractions. In addition, we ensure the local constraint $\phi_a+\phi_c=1$ by introducing a Lagrangian multiplier $\lambda=\lambda(x)$ conjugate to the sum $\phi_a+\phi_c$. This yields for the variation of the free energy
\begin{eqnarray} \label{ht65}
\frac{\delta F}{A}&=&\Psi_0 \frac{\delta \sigma}{e}-\frac{\alpha}{\nu} l^2 \left[\phi_a'(0) \: \delta \phi_a(0)+\phi_c'(0) \: \delta \phi_c(0)\right]\nonumber\\
&+&\frac{1}{\nu} \int \limits_0^\infty dx \: \Big\{ \delta \phi_a \: \left[-\Psi+\ln \phi_a+\chi \phi_c-\alpha l^2 \phi_a''+\lambda\right]\nonumber\\
&+& \delta \phi_c \: \left[\Psi+\ln \phi_c+\chi \phi_a-\alpha l^2 \phi_c''+\lambda\right]\Big\},
\end{eqnarray}
where $\Psi_0=\Psi(x=0)$ denotes the surface potential. The term $\Psi_0 \: \delta \sigma/e$ in the first line of Eq.~\ref{ht65} vanishes because we assume the surface charge density $\sigma$ is fixed. In the absence of specific ion-surface interactions there is no physical reason to fix $\phi_a(0)$ and $\phi_c(0)$. Hence, we adopt the natural boundary conditions $\phi_a'(0)=\phi_c'(0)=0$, which minimize the free energy with respect to the surface values of  $\phi_a$ and $\phi_c$. (Although not considered in the present work, we note that specific ion-surface interactions, or solvent-surface interactions, have been proposed to further increase the ability of ionic liquids to store energy \cite{kondrat16}.) Vanishing of $\delta F$ in thermal equilibrium is finally ensured by the two relations $\phi_a=\exp{(\Psi-\chi \phi_c+\alpha l^2 \phi_a''-\lambda)}$ and $\phi_c=\exp{(-\Psi-\chi \phi_a+\alpha l^2 \phi_c''-\lambda)}$. Choosing $\lambda$ so as to satisfy the constraint $\phi_a+\phi_c=1$ leads to 
\begin{eqnarray} \label{jo94}
\phi_a&=&\frac{1}{1+\exp{\left[-2 \Psi-\chi (\phi_a-\phi_c)-\alpha l^2 (\phi_a''-\phi_c'')\right]}}, \nonumber\\
\phi_c&=&\frac{1}{1+\exp{\left[2 \Psi+\chi (\phi_a-\phi_c)+\alpha l^2 (\phi_a''-\phi_c'')\right]}}.
\end{eqnarray}
These are the equilibrium distribution functions for the anion and cation volume fractions. Eqs.~\ref{jo94} constitute two differential equations from which $\phi_a$ and $\phi_c$ can be calculated for any given value of the potential $\Psi$. Inserting $\phi_a$ and $\phi_c$ from Eq.~\ref{jo94} into the Poisson equation $l^2 \Psi''=\phi_a-\phi_c$ results in the fourth-order nonlinear differential equation
\begin{equation} \label{ni73}
\frac{\alpha}{2} l^4 \Psi'''' +\frac{\chi}{2} l^2 \Psi''+\Psi=\arctanh \left( l^2 \Psi'' \right)
\end{equation}
that must be solved subject to four boundary conditions. The first two, $\Psi(x \rightarrow \infty)=\Psi'(x \rightarrow \infty)=0$, reflect the vanishing of the electric field in the bulk of a structurally stable ionic liquid. The third boundary condition accounts for the fixed surface charge density $\sigma$ at the electrode surface, $x=0$, which we express as $\Psi'(0)=-s/l$, where we have introduced the scaled surface charge density $s=(\nu/l)  \times \sigma/e=4 \pi l_B l \sigma/e$. The fourth boundary condition $\Psi'''(0)=0$ follows from the Poisson equation together with $\phi_a'(0)=\phi_c'(0)=0$. Note that one integration of Eq.~\ref{ni73} can be carried out, leading to 
\begin{eqnarray} \label{jo91}
\frac{\alpha}{4} l^6 \Psi'''^2&=&-\frac{\chi}{4} l^4 \Psi''^2+l^2 \Psi'' \mbox{arctanh} (l^2 \Psi'')\nonumber\\
&+&\frac{1}{2} \ln(1-l^4 \Psi''^2)-l^2 \Psi \Psi''+\frac{1}{2} l^2 \Psi'^2.
\end{eqnarray}
Using our boundary conditions at position $x=0$, this gives rise to the relation
\begin{equation} \label{ju76}
0=-\frac{\chi}{4} c^2+c \: \mbox{arctanh} \: c+\frac{1}{2} \ln(1-c^2)-c \Psi_0+\frac{s^2}{2},
\end{equation}
where here and in the following we use the abbreviation $c=l^2 \Psi''(0)$ for the dimensionless second derivative of the potential at the surface.

Solutions of Eq.~\ref{ni73} (or Eq.~\ref{jo91}) can be used to calculate thermodynamic properties of the EDL. This includes the differential capacitance $C_{diff}=d \sigma/d \Phi(0)$ or, equivalently, the scaled differential capacitance 
\begin{equation} \label{kp01}
\bar{C}_{diff}=\frac{l C_{diff}}{\epsilon \epsilon_0}=\frac{d s}{d \Psi_0}.
\end{equation}
Calculations of $\bar{C}_{diff}$ are based on the relationship between the dimensionless surface potential $\Psi_0$ and the scaled surface charge density $s$.

\section{Results and Discussion}
\subsection{Linear regime, valid for weakly charged electrode}
\begin{figure}[!ht]
\begin{center}
\includegraphics[width=7cm]{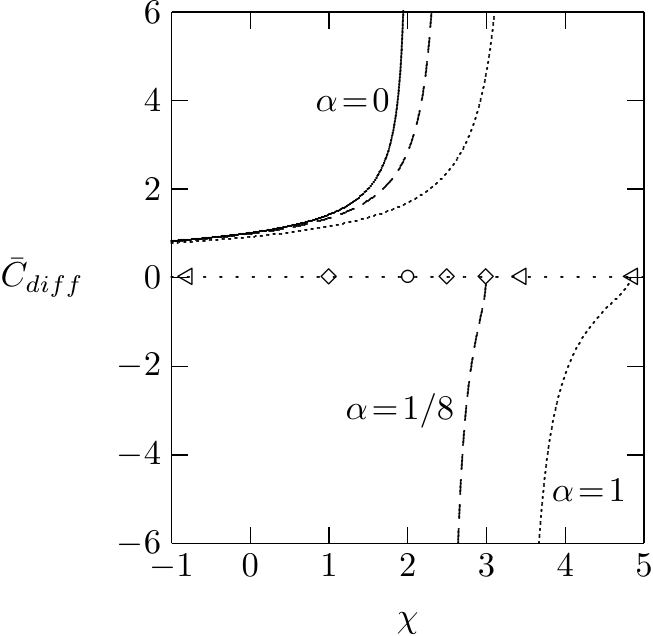}
\caption{\label{fig1}
$\bar{C}_{diff}$ for an uncharged surface ($s=0$) according to Eq.~\ref{ce32} as function of the degree of nonideality $\chi$ for $\alpha=0$ (solid line), $\alpha=1/8$ (dashed lines), and $\alpha=1$ (dotted lines). The locations of $\chi_1$, $\chi_c$, and $\chi_2$ are marked on the center horizontal axis by the symbols ``$\diamond$'' (for $\alpha=1/8$) and ``$\triangleleft$'' (for $\alpha=1$). The symbol ``$\circ$'' marks the position $\chi_1=\chi_c=\chi_2=2$ for $\alpha=0$.
}
\end{center}
\end{figure}

\begin{figure}[!ht]
\begin{center}
\includegraphics[width=7.0cm]{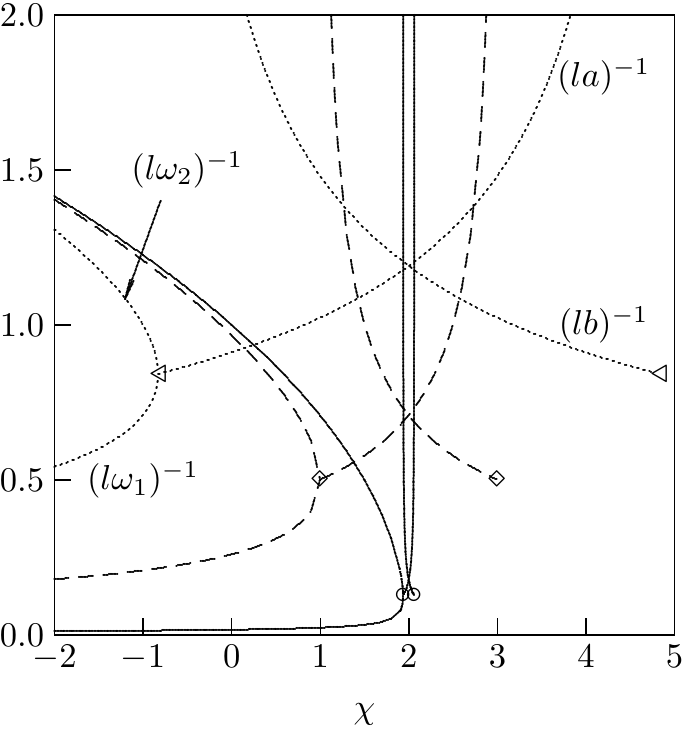}
\caption{\label{fig2}
The scaled characteristic lengths $(l \omega_1)^{-1}$ and $(l \omega_2)^{-1}$ for $\chi < \chi_1$ as well as $(l a)^{-1}$ and $(l b)^{-1}$ for $\chi_1 <\chi <\chi_2$, plotted as function of $\chi$ for different $\alpha$ with $\alpha=1/2000$ (solid lines), $\alpha=1/8$ (dashed lines), and $\alpha=1$ (dotted lines). The four labels $(l \omega_1)^{-1}$, $(l \omega_2)^{-1}$, $(l a)^{-1}$, and $(l b)^{-1}$ mark the four dotted lines; they apply analogously to the sets of dashed and solid lines. The values of $\chi_1$ and $\chi_2$ are marked by the symbol $\triangleleft$ for  $\alpha=1$, by the symbol $\diamond$ for  $\alpha=1/8$, and by the symbol $\circ$ for  $\alpha=1/2000$. Note that $(l b)^{-1}$ diverges at $\chi=\chi_1$ and $(l a)^{-1}$ diverges at $\chi=\chi_2$.}
\end{center}
\end{figure}

We first discuss the limit of small potential, $|\Psi| \ll 1$, which renders Eq.~\ref{ni73} linear, 
\begin{equation} \label{jo62}
\frac{\alpha}{2} l^4 \Psi'''' +\left(\frac{\chi}{2}-1\right) l^2 \Psi''+\Psi=0.
\end{equation}
The fourth-order nature of the linearized mean-field equation for the potential is consistent with the phenomenological model proposed by Bazant {\em et al} \cite{bazant11} and with the general approach by Blossey {\em et al} \cite{blossey17}. Eq.~\ref{jo62} gives rise to the two length scales, $1/\omega_1$ and $1/\omega_2$, that follow from the characteristic equation $(\alpha/2) \: l^4 \omega^4 +[(\chi/2)-1] \: l^2 \omega^2+1=0$. We find
\begin{equation} \label{hu51}
(l \omega_{1/2})^2=\frac{1}{\alpha} \left\{\left(1-\frac{\chi}{2}\right) \pm \sqrt{\left(1-\frac{\chi}{2}\right)^2-2 \alpha}\right\}.
\end{equation}
The potential that satisfies Eq.~\ref{jo62} subject to $\Psi'(0)=-s/l$ and $\Psi'''(0)=\Psi(x \rightarrow \infty)=\Psi'(x \rightarrow \infty)=0$ can conveniently be expressed in terms of $\omega_1$ and $\omega_2$,
\begin{equation} \label{gy65}
\Psi(x)=\frac{s}{l} \: \frac{\omega_1^3 e^{-\omega_2 x}-\omega_2^3 e^{-\omega_1 x}}{\omega_1^3 \omega_2-\omega_1 \omega_2^3},
\end{equation}
which implies for the differential capacitance
\begin{equation} \label{ce32}
  \bar{C}_{diff}=\frac{l}{\frac{1}{\omega_1}+\frac{1}{\omega_2}-\frac{1}{\omega_1+\omega_2}}=\frac{\sqrt{1-\frac{\chi}{2}+\sqrt{2 \alpha}}}{1-\frac{\chi}{2}+\frac{1}{2} \sqrt{2 \alpha}}.
\end{equation}
Eq.~\ref{ce32} is a major result of the present work. Fig.~\ref{fig1} shows $\bar{C}_{diff}$ as function of $\chi$ for three different values of $\alpha$.
Before we discuss Eq.~\ref{ce32} and Fig.~\ref{fig1} it is useful to introduce three special values for $\chi$ that we denote by $\chi_1$, $\chi_2$, and $\chi_c$. The first two, $\chi_1$ and $\chi_2$, specify the solutions
\begin{equation} \label{hu73}
\chi_1=2 \left(1-\sqrt{2 \alpha}\right), \hspace{1cm} 
\chi_2=2 \left(1+\sqrt{2 \alpha}\right)
\end{equation}
of the equation $(1-\chi/2)^2=2 \alpha$ at which the square root in Eq.~\ref{hu51} vanishes. The third one is defined through
\begin{equation} \label{ji27}
\chi_c=2 \left(1+\frac{1}{2} \sqrt{2 \alpha}\right).
\end{equation}
For $\chi<\chi_1$ the two inverse lengths $\omega_1$ and $\omega_2$ are both positive real numbers, implying that the potential $\Psi(x)$ is the sum of two exponential functions and thus is monotonically decreasing with growing $x$. For $\chi_1<\chi<\chi_2$ both $\omega_1=a-i b$ and $\omega_2=a+i b$ have nonvanishing real and imaginary parts
\begin{equation} \label{fe12}
a=\frac{1}{l} \sqrt{\frac{1}{\sqrt{2 \alpha}}-\frac{\chi/2-1}{2 \alpha}}, \hspace{0.5cm}
b=\frac{1}{l} \sqrt{\frac{1}{\sqrt{2 \alpha}}+\frac{\chi/2-1}{2 \alpha}},
\end{equation} 
and the potential $\Psi(x)$ exhibits damped oscillations. Indeed, expressing the potential in Eq.~\ref{gy65} explicitly in terms of the inverse lengths $a$ and $b$,
\begin{equation} \label{du58}
\Psi(x)=\frac{s}{l} \: \frac{e^{-a x}}{a^2+b^2} \left[\frac{3 a^2-b^2}{2 a} \cos( b x)-\frac{3 b^2-a^2}{2 b} \sin( b x)\right],
\end{equation} 
clearly reveals its spatially decaying oscillations. Fig.~\ref{fig2} shows a plot of the two characteristic lengths, $1/\omega_1$ and $1/\omega_2$ for $\chi<\chi_1$ as well as $1/a$ and $1/b$ for $\chi_1<\chi<\chi_2$, as function of $\chi$. 
Note that every value of $\chi$, apart from $\chi=\chi_1$, $\chi=2$, and $\chi=\chi_2$, is associated with two distinct characteristic lengths and that all displayed lengths in Fig.~\ref{fig2} are scaled by $l$. Our model predicts a large decay length in the regime of a non-oscillating electrostatic potential (as in Eq.~\ref{gy65}) for very negative $\chi$ and in the regime of an oscillating potential (as in Eq.~\ref{du58}) when $\chi$ is positive and approaches $\chi_2$. Large decay lengths in highly concentrated electrolytes have been measured \cite{gebbie17,perez17} and are being discussed in the framework of ``underscreening''.

For $\chi>\chi_2$ the real parts but not the imaginary parts of $\omega_1$ and $\omega_2$ in Eq.~\ref{hu51} vanish, implying $\Psi(x)$ is oscillating, with two characteristic length scales. The ionic liquid is structurally unstable in this case. However, macroscopic phase separation is prohibited due to the excess charge of each phase. As a result, a doubly modulated phase with two characteristic inverse length scales $\tilde{a}=-i a$ and $b$, where $a$ and $b$ are specified in Eq.~\ref{fe12}, appears in thermal equilibrium. The nonvanishing potential of the bulk phase is then of the form $\Psi(x) \sim \sin(\tilde{a} x) \cos(b x)$. A systematic analysis of the equilibrium structure of a bulk ionic liquid in three dimensional space is outside the scope of the present study. Hence, in all this work we assume $\chi<\chi_2$. Recall that we also assume $\alpha \ge 0$ because negative values of $\alpha$ lead to a structurally unstable ionic liquid for any choice of $\chi$.

After having introduced $\chi_1$, $\chi_2$, and $\chi_c$, we are now in a position to discuss the differential capacitance $\bar{C}_{diff}$ as specified in Eq.~\ref{ce32} and plotted in Fig.~\ref{fig1}. For growing $\chi$ the differential capacitance increases, with no discontinuities when $\chi$ passes through $\chi_1$, where the potential begins to exhibit damped oscillations. When $\chi$ approaches $\chi_c$, the differential capacitance diverges; $\bar{C}_{diff} \rightarrow \infty$. Upon crossing $\chi_c$ the sign of $\bar{C}_{diff}$ flips. Subsequently, in the region  $\chi_c<\chi<\chi_2$, the differential capacitance grows from large negative values to zero. This behavior is qualitatively the same for any choice of $\alpha>0$, and is displayed in Fig.~\ref{fig1} for $\alpha=0$ (solid line), $\alpha=1/8$ (dashed line), and $\alpha=1$ (dotted line). The three points $\chi_1$, $\chi_c$, and $\chi_2$ are marked on the center horizontal axis by the three symbols ``$\diamond$'' for $\alpha=1/8$ and by the three symbols ``$\triangleleft$'' for $\alpha=1$. For $\alpha=0$ the three points $\chi_1=\chi_c=\chi_2=2$ (marked by the symbol ``$\circ$'' on the center horizontal axis) are degenerate.

\begin{figure}[!ht]
\begin{center}
\includegraphics[width=8cm]{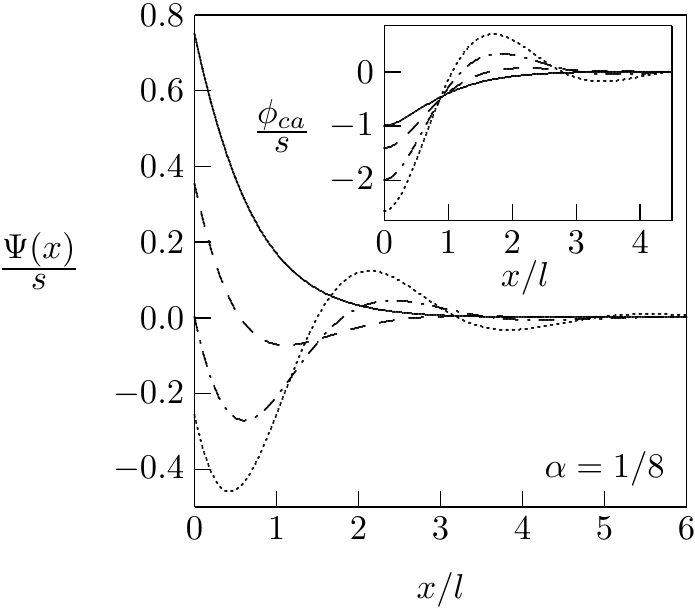}
\caption{\label{fig3}
$\Psi(x)/s$ as function of the scaled distance $x/l$ according to Eq.~\ref{du58} for $\alpha=1/8$ and $\chi=\chi_1=1.0$ (solid line), $\chi=2.0$ (dashed line), $\chi=\chi_c=2.5$ (dash-dotted line), and $\chi=2.7$ (dotted line). The inset shows the corresponding difference $\phi_{ca}=\phi_c-\phi_a$ of the volume fraction of cations and anions, calculated according to Eq.~\ref{sr45} and scaled by $s$.
}
\end{center}
\end{figure}

The divergence of $\bar{C}_{diff}$ for $\chi=\chi_c$ and its negative sign in the region $\chi_c<\chi<\chi_2$ are notable. To understand its physical origin, we plot in Fig.~\ref{fig3} the scaled potential $\Psi(x)/s$ for $\alpha=1/8$ and the four different choices $\chi=\chi_1=1$ (solid line), $\chi=2$ (dashed line), $\chi=\chi_c=2.5$ (dash-dotted line), and $\chi=2.7$ (dotted line). 
Recall that for $\chi=\chi_2=3$ (this case is not displayed in Fig.~\ref{fig3}) the electrolyte becomes structurally unstable, exhibiting oscillations of $\Psi(x)$ that do not decay for large $x$. For $\chi=\chi_c$ the surface potential vanishes, $\Psi_0=0$. That is, the surface potential remains zero, even when equipping the surface with a charge density as expressed by the nonvanishing slope $\Psi'(0)=-s/l$; see the dash-dotted line in Fig.~\ref{fig3}. Hence, the differential capacitance $\bar{C}_{diff}$ must be infinitely high. The vanishing of the surface potential $\Psi(0)=0$ at $\chi=\chi_c$ is a general property for any choice of $\alpha>0$. This can be demonstrated mathematically by inserting $\chi_c$ into $a$ and $b$ and that into $\Psi(0)$ according to Eq.~\ref{du58}. When $\chi$ is further increased beyond $\chi_c$ (while maintaining structural stability, $\chi<\chi_2$), the surface potential of a positively charged surface adopts negative values, $\Psi(0)<0$, and small changes of $ds$ and $d\Psi_0$ carry different signs. This implies a negative sign of the differential capacitance in the entire region $\chi_c<\chi<\chi_2$. As $\chi$ approaches $\chi_2$ the scaled surface potential $\Psi(0)/s$ grows to large negative values thus rendering the magnitude of $\bar{C}_{diff}$ small.

In the inset of Fig.~\ref{fig3} we show the difference $\phi_{ca}=\phi_c-\phi_a$ of the volume fraction of cations and anions that corresponds to the four cases displayed in the main diagram of Fig.~\ref{fig3}. We point out that $\phi_{ca}$ is proportional to the local charge density; $\phi_{ca}>0$ implies a local excess of cations and $\phi_{ca}<0$ a local excess of anions. We calculate $\phi_{ca}$ as the solution of the inhomogeneous linear differential equation 
\begin{equation} \label{jk80}
\frac{\alpha}{2} l^2 \phi_{ca}''-\left(1-\frac{\chi}{2}\right) \phi_{ca}=\Psi,
\end{equation} 
subject to the boundary conditions $\phi_{ca}'(x=0)=0$ and $\phi_{ca}(x \rightarrow \infty)=0$. Note that Eq.~\ref{jk80} follows from linearizing the distribution functions for $\phi_a$ and $\phi_c$ in Eq.~\ref{jo94}. The potential $\Psi(x)$ that enters Eq.~\ref{jk80} is specified explicitly in Eq.~\ref{du58}. The solution of Eq.~\ref{jk80} can be written as
\begin{equation} \label{sr45}
\phi_{ca}=-\frac{s}{l} \times \frac{e^{-a x}}{\sqrt{2 \alpha}} \: 
\left[\frac{\cos(b x)}{a}+\frac{\sin(b x)}{b} \right],
\end{equation} 
where we recall the definitions of the inverse length scales $a$ and $b$ in Eq.~\ref{fe12}. Of course, knowing $\phi_{ca}$ gives us immediate access to $\phi_a=(1-\phi_{ca})/2$ and $\phi_c=(1+\phi_{ca})/2$. The damped oscillations of $\phi_{ca}$ predicted by Eq.~\ref{sr45} and visualized in the inset of Fig.~\ref{fig3} become more pronounced when $\chi$ approaches $\chi_2$. This also follows from our preceding analysis of the characteristic length $1/a$; see Fig.~\ref{fig2}. The presence of layered domains alternatingly enriched in anions and cations together with the growing characteristic length $1/a$ embodies the concepts of ``overscreening'' and ``underscreening'' that are being used to rationalize the structure of the EDL for ionic liquids \cite{bazant11,perez17,gebbie17}. 

We proceed with a physical interpretation of our predictions, especially the formation of layered domains with excess anions and cations that can even lead to a negative differential capacitance. Clearly, growing $\chi$ increases the tendency of like-charged ions to cluster. We point out that electrostatic short-range interactions act towards rendering  $\chi$ negative. Our model is also applicable to $\chi<0$, but an oscillating potential is not predicted in this case. The physically most interesting regime is that of positive $\chi$, with $\chi_1 < \chi <\chi_2$. We do not know what physical mechanism could cause $\chi$ to become positive, even less so to approach $\chi_2$ where the ionic liquid becomes intrinsically unstable. However, ions that form ionic liquids are typically bulky, often with short hydrocarbon side chains or other moieties that trigger hydrophobic interactions. These non-electrostatic interactions can tune the value of $\chi$ and possibly also its sign. In fact, non-electrostatic attraction between like-charged ions can be expected to be more abundant than repulsion because the competition between non-electrostatic attraction and Coulomb repulsion acts toward stabilizing the fluid state of an ionic liquid. For $\chi>\chi_1$ the tendency of like-charged ions to cluster leads to an over-neutralization of the surface charges by the first layer of counterions. At the same time, because of that additional enrichment of counterions close to the surface, the scaled surface potential $\Psi_0/s$ decreases. As $\chi$  grows further, the formation of alternating layers enriched in counterions and coions becomes more pronounced (see the inset of Fig.~\ref{fig3}) and at $\chi=\chi_c$, the surface potential $\Psi_0=0$ vanishes (see the dash-dotted line in Fig.~\ref{fig3}). The vanishing surface potential for non-vanishing surface charge density is a consequence of the strongly nonideal behavior of the ionic liquid: the number of attracted counterions overcharges the surface in such a manner that the two potential contributions originating from the surface and from  the counterions exactly cancel each other. Further beyond this inversion point, in the region $\chi_c < \chi < \chi_2$, a positive surface charge density induces a negative surface potential (and vice versa). These considerations, shown in Fig.~\ref{fig3} for $\alpha=1/8$ (see the dotted lines), are general and apply to any choice of $\alpha$ (with $\alpha>0$). The possible enrichment of cations close to an electrode with a positive surface potential -- which is a prediction of our present model -- has indeed been reported experimentally \cite{lauw12} and was rationalized by the presence of specific interactions between the electrode surface and the ions of the ionic liquid. Our present model suggests that, in principle, no such specific interactions are needed if the ionic liquid exhibits sufficiently strong nonideal behavior.

\subsection{Non-linear regime, valid for any electrode charge}
\begin{figure}[!ht]
\begin{center}
\includegraphics[width=7cm]{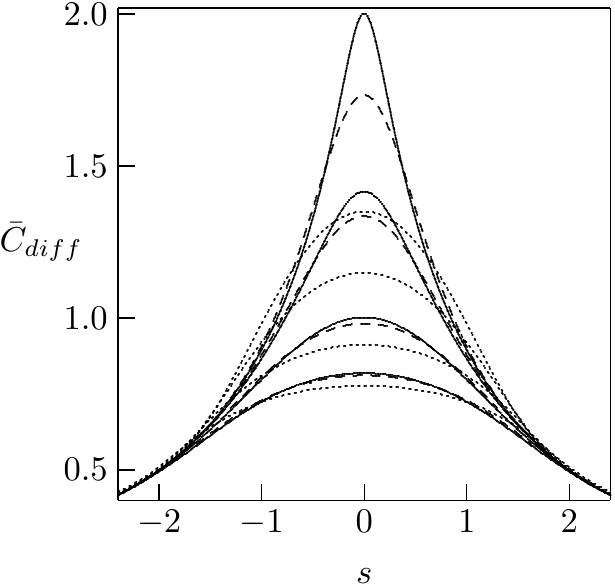}
\caption{\label{fig4}
$\bar{C}_{diff}$ as function of $s$ for $\alpha=0$ (solid lines), $\alpha=1/8$ (dashed lines), and $\alpha=1$ (dotted lines), in each case for $\chi=1.5$, $\chi=1$, $\chi=0$, and $\chi=-1$ (from top to bottom). 
}
\end{center}
\end{figure}

In the remainder of this work we study the full nonlinear problem, based on Eq.~\ref{ni73}. To compute numerical solutions of Eq.~\ref{ni73} we adopt a Newton-Raphson iteration scheme
\begin{eqnarray} \label{vt67}
\frac{\alpha}{2} l^4 \Psi_{i+1}''''&+&\left(\frac{\chi}{2}-\frac{1}{1-l^4 \Psi_i''^2}\right) l^2 \Psi_{i+1}''+\Psi_{i+1}\nonumber\\
&=&-\frac{l^2 \Psi_i''}{1-l^4 \Psi_i''^2}+{\rm arctanh} (l^2 \Psi_i''),
\end{eqnarray}
subject to the boundary conditions $\Psi_{i+1}'(0)=-s/l$ and $\Psi_{i+1}'''(0)=\Psi_{i+1}(x \rightarrow \infty)=\Psi_{i+1}'(x \rightarrow \infty)=0$ for all non-negative integers $i$ starting from $i=0$. When convergence is reached ($\Psi_{i+1}=\Psi_i$), the solution satisfies Eq.~\ref{ni73}. From the numerically given relation $\Psi_0=\Psi_0(s)$ we calculate $\bar{C}_{diff}=\bar{C}_{diff}(s,\chi,\alpha)$ directly through $\bar{C}_{diff}=(d \Psi_0/d s)^{-1}$. Fig.~\ref{fig4} presents $\bar{C}_{diff}$ as function of $s$ for $\alpha=0$ (solid lines), $\alpha=1/8$ (dashed lines), and $\alpha=1$ (dotted lines).
The four curves for each $\alpha$ correspond to $\chi=1.5$, $\chi=1$, $\chi=0$, and $\chi=-1$ (from top to bottom). In the absence of nearest-neighbor interactions, for $\chi=\alpha=0$, the differential capacitance is given by \cite{kornyshev07} $\bar{C}_{diff}=\sqrt{1-e^{-s^2}}/|s|$. A perturbation calculation for small $\chi$ (still with $\alpha=0$) yields 
\begin{equation} \label{kl20}
\bar{C}_{diff}=\frac{1}{|s|} \sqrt{1-e^{-s^2}} \left(1+ \frac{\chi}{4} e^{-s^2}\right),
\end{equation}
which provides a good approximation for $|\chi| \lesssim 1/2$. When expressed as function of the surface potential $\Psi_0$, the differential capacitance for small $\chi$ (and $\alpha=0$) reads
\begin{equation} \label{vv54}
\bar{C}_{diff}=\frac{|\tanh \Psi_0 |}{\sqrt{2 \ln(\cosh \Psi_0)}} \left[1+\chi
\frac{\frac{1+4 \ln(\cosh \Psi_0)}{\cosh^2 \Psi_0 }-1}{8 \ln(\cosh \Psi_0)}
\right].
\end{equation}
Note that, when expressed as function of $\Psi_0$, our results for $\alpha=0$ coincide with the results of Goodwin {\em et al} \cite{goodwin17}. We also point out the order of magnitude $\bar{C}_{diff}(s) \approx 1$ for all values displayed in Fig.~\ref{fig4}. The estimate $\bar{C}_{diff}(s)=1$ is equivalent to $C_{diff}=\epsilon \epsilon_0/l$ (see Eq.~\ref{kp01}), where we recall $l=\sqrt{\nu/(4 \pi l_B)}$ with the Bjerrum length $l_B=e^2/(4 \pi \epsilon \epsilon_0 k_BT)$. Based on a dielectric constant of $\epsilon=5$, an ion volume $\nu=1 \: \mbox{nm}^3$, and room temperature ($T=300 \: \mbox{K}$), we find $C_{diff}=e \sqrt{\epsilon \epsilon_0/(\nu k_BT)}=0.5 \: \mbox{F}/\mbox{m}^2$. This compares well with typical experimental values \cite{seddon18}, which are on the order of $C_{diff}=0.1 \: \mbox{F}/\mbox{m}^2$.

All values for $\bar{C}_{diff}$ calculated at $s=0$ in Fig.~\ref{fig4} follow the predictions of Eq.~\ref{ce32}. We observe that for small $|s|$ the differential capacitance $\bar{C}_{diff}$ increases as $\chi$ becomes more positive. This is physically plausible because more attractive nearest-neighbor interactions between ions of the same type tend to more strongly condense the counterion layer that forms near the electrode surface. If the EDL is represented by a parallel-plate capacitor with an effective distance $d$ between the two plates, then a more condensed counterion layer entails a smaller $d$ and thus a larger $\bar{C}_{diff}$. This mechanism ceases to apply to highly charged electrodes (with $|s| \gg 1$) because in that case the counterions are already condensed, with virtually no coions present that could be expelled by more attractive counterion-counterion interactions. Therefore, for $|s| \gg 1$ the differential capacitance remains largely unaffected by $\chi$ and $\alpha$. In fact, as pointed out by Kilic {\em et al} \cite{kilic07}, for $|s| \gg 1$ all curves merge into the single behavior $\bar{C}_{diff}=1/|s|$. Growing values of $\alpha$ widen the bell-shaped curves of $\bar{C}_{diff}$; they decrease for sufficiently small $|s|$ and slightly increase for intermediate $|s|$.

\begin{figure}[!ht]
\begin{center}
\includegraphics[width=7.5cm]{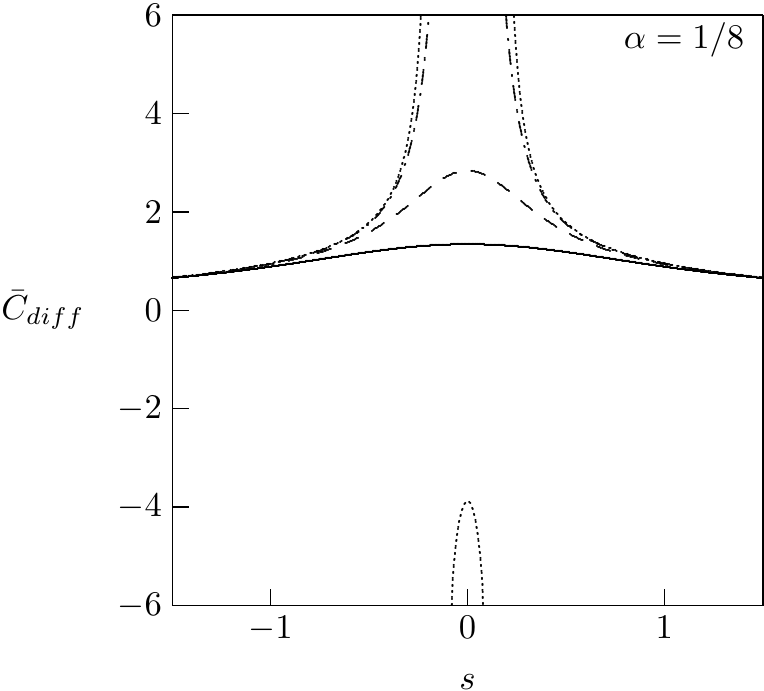}
\caption{\label{fig5}
$\bar{C}_{diff}$ as function of $s$ for $\alpha=1/8$. The different curves correspond to $\chi=\chi_1=1.0$ (solid line), $\chi=2.0$ (dashed line), $\chi=\chi_c=2.5$ (dash-dotted line), and $\chi=2.7$ (dotted line).
}
\end{center}
\end{figure}

\begin{figure}[!ht]
\begin{center}
\includegraphics[width=8cm]{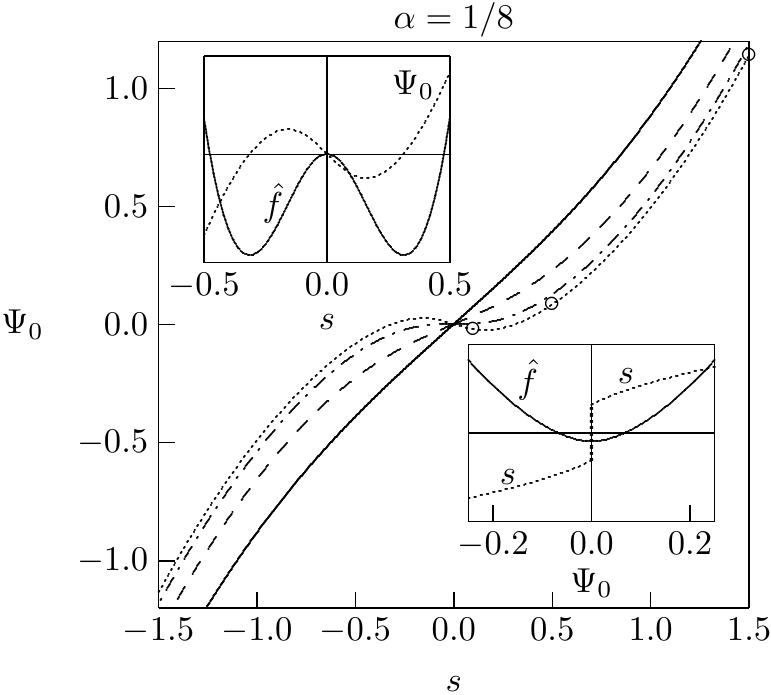}
\caption{\label{fig6}
Surface potential $\Psi_0$ as function of $s$ for $\alpha=1/8$. The different curves correspond to $\chi=\chi_1=1.0$ (solid line), $\chi=2.0$ (dashed line), $\chi=\chi_c=2.5$ (dash-dotted line), and $\chi=2.7$ (dotted line). The three circular symbols $\circ$ mark the three cases for which the potential $\Psi(x)$ and volume fraction $\phi_a(x)$ of the counterions are displayed in Fig.~\ref{fig7}. The upper left inset re-displays $\Psi_0(s)$ (dotted line) and the corresponding free energy $\hat{f}(s)=\int_0^s d\bar{s} \: \Psi_0(\bar{s})$ (solid line) for $\chi=2.7$. The lower right inset re-displays the stable branch of the inverse function $s(\Psi_0)$ (dotted line) and the corresponding free energy $\hat{f}[s(\Psi_0)]=\int_0^{\Psi_0} d\bar{\Psi}_0 \: \bar{\Psi}_0 \: \:\bar{C}_{diff}(\bar{\Psi}_0)$ (solid line) for $\chi=2.7$. To improve visibility, we have magnified $\hat{f}$ in both insets  by a factor of 20.
}
\end{center}
\end{figure}

All examples shown in Fig.~\ref{fig4} are characterized by $\chi<\chi_c$, where the differential capacitance $\bar{C}_{diff}(s=0)$ does not diverge. We also discuss the nonlinear problem for $\chi=\chi_c$ and for $\chi_c<\chi<\chi_2$. To this end, Fig.~\ref{fig5} shows the differential capacitance $\bar{C}_{diff}$ for $\alpha=1/8$ for the same choices of $\chi$ as in Fig.~\ref{fig3}.
The curve in Fig.~\ref{fig5} for $\chi=\chi_1=1$ (see the solid line) has already been displayed as one of the curves in Fig.~\ref{fig4}. At $\chi=\chi_c=2.5$ we observe $\bar{C}_{diff}$ to diverge at $s=0$ as predicted by Eq.~\ref{ce32}. For $\chi=2.7$ (which is located between $\chi_c=2.5$ and $\chi_2=3$), the differential capacitance does again diverge, yet at two nonvanishing values $\pm|s|$. Between these two values, $\bar{C}_{diff}$ adopts negative values, with a minimal magnitude at $s=0$; see the dotted line in the lower half of Fig.~\ref{fig5}. The minimum value $\bar{C}_{diff}=-3.873$ at $s=0$ agrees, of course, with the prediction of Eq.~\ref{ce32}. In order to better understand the behavior in Fig.~\ref{fig5}, we display in Fig.~\ref{fig6} the surface potential as function of the surface charge density, $\Psi_0(s)$, for the same values of $\alpha$ and $\chi$ as in Fig.~\ref{fig5}.
For $\chi=\chi_c=2.5$ (see the dash-dotted line in Fig.~\ref{fig6}) the slope of $\Psi_0(s)$ at $s=0$ vanishes and this leads to the diverging differential capacitance at $s=0$ as predicted by Eq.~\ref{ce32}. For $\chi=2.7$ (see the dotted line in Fig.~\ref{fig6}) the surface potential $\Psi_0$ initially decreases to negative values as a response to a small positive surface charge density ($s>0$). Only for larger values of $s$, the surface potential starts to increase and eventually adopts positive values. The vanishing slope of $\Psi_0(s)$ at the nonvanishing values of $\pm|s|$ triggers $\bar{C}_{diff}$ to diverge. In between, from $-|s|$ to $+|s|$, the negative slope of $\Psi_0(s)$ implies the differential capacitance is negative; see the dotted line in the lower half of Fig.~\ref{fig5}.

Given the negative slope of $\Psi_0(s)$ for $\chi>\chi_c$ and small $s$, it is interesting to investigate the stability of the EDL in that regime. To this end, we consider the charging free energy (per unit area) of the EDL
\begin{equation} \label{ji26}
\frac{F}{A}=\frac{l}{\nu} \int \limits_0^s d\bar{s} \: \Psi_0(\bar{s})=\frac{l}{\nu} \int \limits_0^{\Psi_0} d\bar{\Psi}_0 \: \bar{\Psi}_0 \: \:\bar{C}_{diff}(\bar{\Psi}_0),
\end{equation}
which follows from the free energy variation in Eq.~\ref{ht65} and from the definition of the scaled differential capacitance in Eq.~\ref{kp01}. The self-consistency relation in Eq.~\ref{ni73} must be fulfilled during the charging process. The free energy in Eq.~\ref{ji26} is the excess free energy of the EDL with respect to the bulk (it is identical to $F/A$ defined in Eq.~\ref{sw31} subject to a constant that selects the bulk as reference). 
The two insets in Fig.~\ref{fig6} display the scaled free energy $\hat{f}=F \nu/(A l)$ for the case $\alpha=1/8$ and $\chi=2.7$. The upper left inset in Fig.~\ref{fig6} re-displays $\Psi_0(s)$ together with its integral $\hat{f}(s)=\int_0^s d\bar{s} \: \Psi_0(\bar{s})$. In this case, the charge density $s$ on the electrode is fixed. The charging free energy is negative for sufficiently small $|s|$, adopting a minimum at $s=\pm s^\star$ where $\Psi_0(\pm s^\star)=0$. Although the EDL adopts a negative free energy within the region $-s^\star < s < s^\star$, the system is stable because $s$ is fixed and the EDL has no additional degrees of freedom. Any ability of the surface to laterally redistribute charges leads to an instability \cite{partenskii08,partenskii11}. This is most pronounced when the surface potential $\Psi_0$ is fixed. For example, demanding $\Psi_0=0$ leaves the three choices $s=0$ and $s=\pm s^\star$ for the surface charge density. The free energy for $s=\pm s^\star$ is lower than that for $s=0$. In fact $\hat{f}(\pm s^\star)$ is the smallest possible value that the free energy can adopt; see the upper left inset in Fig.~\ref{fig6}. Hence, the EDL will spontaneously adopt a non-vanishing charge $s=\pm s^\star$ despite its vanishing surface potential. The increase of the surface potential to a small positive value $\Psi_0$ selects the largest of the three solutions for $s$ of the equation $\Psi_0=\Psi_0(s)$, a solution with $s>s^{\star}$ and an energy that is still negative but larger than that for $\Psi_0=0$. Even more positive $\Psi_0$ eventually leads to $\hat{f}(s)>0$ and only one single solution of the equation $\Psi_0=\Psi_0(s)$. The behavior of $\hat{f}[s(\Psi_0)]=\int_0^{\Psi_0} d\bar{\Psi}_0 \: \bar{\Psi}_0 \: \:\bar{C}_{diff}(\bar{\Psi}_0)$, together with the stable branch of $s(\Psi_0)$ (with the jump from $-s^\star$ to $+s^\star$ at $\Psi_0=0$), is shown in the lower right inset in Fig.~\ref{fig6}. The discontinuous change in $s$ at $\Psi_0=0$ can also be deduced by considering the thermodynamic potential $\check{f}(\Psi_0)=\hat{f}[s(\Psi_0)]-\Psi_0 \: s(\Psi_0)$ (not shown in Fig.~\ref{fig6}). We emphasize that the two insets of Fig.~\ref{fig6} refer to $\chi>\chi_c$. No lowering of the free energy $\hat{f}$ below zero is possible for $\chi<\chi_c$.

The most notable observation from the two insets of Fig.~\ref{fig6} is that for $\chi>\chi_c$ the free energy $\hat{f}$ can adopt negative values. Since $\hat{f}$ is measured with respect to the bulk, the gain in free energy must be provided by the ionic liquid. In other words, the ionic liquid stores a {\em frustration energy} that it is able to release upon contact with the electrode. For example, when a grounded electrode (with $\Psi_0=0$) makes contact with the ionic liquid it spontaneously adopts a positive or negative surface charge and lowers the free energy of the ionic liquid. Small changes of the surface potential from slightly negative to positive values flip the sign of the adopted surface charge. We can draw a direct analogy to a second order transition in thermodynamics; the free energy $\hat{f}(s)$ in the upper left inset of Fig.~\ref{fig6} can be viewed as a Landau free energy below the critical temperature, and $\chi=\chi_c$ characterizes the critical point \cite{partenskii08}. The frustration energy in the ionic liquid arises from the interplay between the nearest-neighbor interactions and electrostatic interactions. The former favor and the latter oppose phase separation between the two ionic species. Inserting an electrode that has (or is allowed to adopt) a small but non-vanishing surface charge density enables the ionic liquid to undergo a local microscopic phase separation, which spatially decays into the bulk; see the inset of Fig.~\ref{fig3}. Within its bulk, the ionic liquid remains stable with respect to spontaneous internal microdomain formation or structural disintegration. It only has a high affinity for surfaces that either carry a small charge or are able to acquire a small charge. Negative differential capacitance of an EDL has been predicted occasionally in previous works based on non-local density functional theories and computer simulations \cite{torrie92,boda02,boda04,gonzalez04,guerrero05}. The corresponding thermodynamic behavior of the EDL has been analyzed by Partenskii and Jordan  \cite{partenskii05,partenskii08,partenskii11} based on general arguments and using the instructive analytic textbook example of a squishy capacitor.

\begin{figure}[!ht]
\begin{center}
\includegraphics[width=8cm]{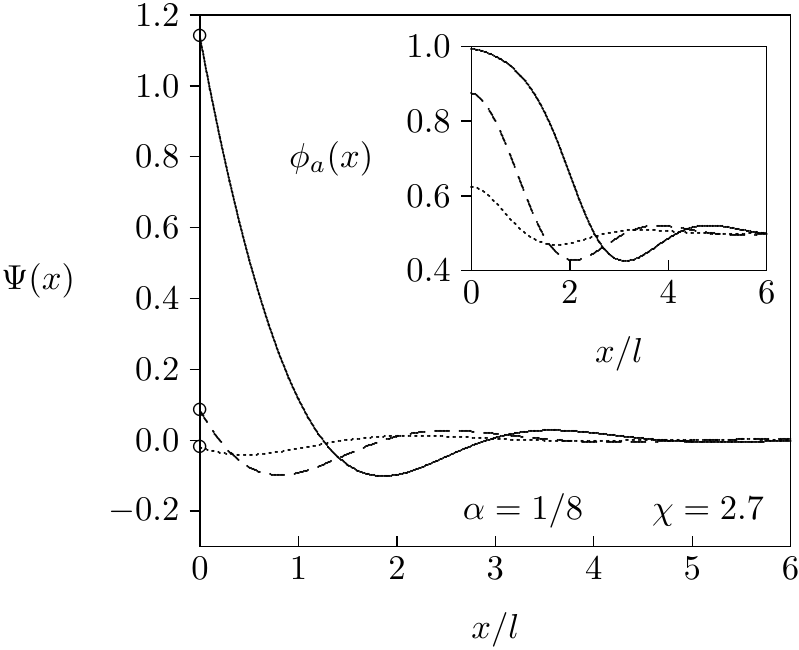}
\caption{\label{fig7} Potential $\Psi(x)$ as function of the scaled distance $x/l$ for $\alpha=1/8$ and $\chi=2.7$. The different curves correspond to $s=0.1$ (dotted line), $s=0.5$ (dashed line), and $s=1.5$ (solid line). The surface potentials $\Psi_0=\Psi(x=0)$ are marked by the symbols $\circ$; they match the symbols $\circ$ in Fig.~\ref{fig6} that are located on the curve for $\chi=2.7$. The inset shows the corresponding volume fraction of the anions, $\phi_a(x)$, as function of $x/l$.}
\end{center}
\end{figure}

In Fig.~\ref{fig7} we investigate how the structure of the EDL adjusts as the surface charge density on the electrode is increased. 
The three plotted cases all refer to $\alpha=1/8$ and $\chi=2.7$, with a scaled surface charge density $s=0.1$, $s=0.5$, and $s=1.5$. The corresponding surface potentials are marked in Figs.~\ref{fig6} and \ref{fig7} by the symbol $\circ$. Clearly, when $s$ increases, the damped oscillations of both potential and counterion density persist even after counterion saturation ($\phi_a \rightarrow 1$) sets in. This prediction is similar to that drawn by Bazant {\em et al} \cite{bazant11} using their phenomenological model that includes short-ranged ion-ion correlations. In the present model, however, an effective short-range attraction between like-charged ions is needed to produce an oscillating potential.

We point out again that the unusual behavior of $\Psi_0(s)$ (and thus of $\bar{C}_{diff}$) is a result of the competition between two different length scales that emerge from the nonvanishing $\alpha$ and $\chi$. No such behavior is observed if $\alpha=0$ is assumed. However, the case $\alpha=0$ allows for the calculation of an implicit relation for $\bar{C}_{diff}$ in the nonlinear theory, as we demonstrate in the following. When applied to the position $x=0$, Eq.~\ref{ni73} with $\alpha=0$ gives rise to 
\begin{equation} \label{hy66}
\Psi_0=\arctanh(c)-\frac{\chi}{2} c,
\end{equation}
where we recall the abbreviation $c=l^2 \Psi''(0)$ for the dimensionless second derivative of the surface potential (initially introduced in Eq.~\ref{ju76}). Inserting Eq.~\ref{hy66} into Eq.~\ref{ju76} yields
\begin{equation} \label{do58}
s=\sqrt{-\ln (1-c^2)-\frac{\chi}{2} c^2}.
\end{equation}
The two relations, $s$ as function of $c$ in Eq.~\ref{do58} and $\Psi_0$ as function of $c$ in Eq.~\ref{hy66} can now be used to calculate the scaled differential capacitance
\begin{equation} \label{yu67}
\bar{C}_{diff}=\frac{d s}{d \Psi_0}=\frac{\frac{d s}{d c}}{\frac{d \Psi_0}{d c}}=\frac{c}{s}
\end{equation}
and therefore
\begin{equation} \label{ji81}
\frac{1}{\bar{C}_{diff}}=\sqrt{-\frac{\ln(1-c^2)}{c^2}-\frac{\chi}{2}}.
\end{equation}
Eq.~\ref{ji81} expresses the differential capacitance exclusively as function of $c$. Using Eqs.~\ref{hy66} and \ref{ji81} we can compute the scaled differential capacitance $\bar{C}_{diff}$ as function of the surface potential $\Psi_0$. Similarly, Eqs.~\ref{do58} and \ref{ji81} give us access to the scaled differential capacitance $\bar{C}_{diff}$ as function of the scaled surface charge density $s$. Recall that Fig.~\ref{fig4} includes plots of $\bar{C}_{diff}(s)$ for $\alpha=0$ (the four solid lines in Fig.~\ref{fig4} were calculated using Eqs.~\ref{do58} and \ref{ji81}). Also, the perturbation results in Eqs.~\ref{kl20} and \ref{vv54} were obtained using Eqs.~\ref{hy66}, \ref{do58}, and \ref{ji81}, with the assumption of $|\chi|\ll 1$.

\section{Conclusions}
The inclusion of nearest-neighbor interactions into the lattice gas model of an incompressible ionic liquid leads to distinct regimes for the behavior of the electrostatic potential and the ensuing differential capacitance. If the ionic liquid exhibits strongly nonideal properties, close to a structural instability, the differential capacitance is predicted to transition through a divergence or to even adopt negative values. At the same time, the EDL is characterized by the formation of layers of alternating excess charge close to the electrode that induce the surface potential to change nonmonotonically as function of the surface charge density. Our present mean-field treatment includes nearest-neighbor interactions subject to two assumptions: no specific ion-surface interactions are present, and anion-anion interactions are the same as cation-cation interactions ($\omega_{aa}=\omega_{cc}$). It is the latter assumption that allowed us to express the self-consistency relation (Eq.~\ref{ni73}) exclusively in terms of the potential $\Psi$. Future work may extend the present model beyond these two restrictions.

We finally point out that the physically most interesting case emerging from our model pertains to $\alpha>0$ and $\chi>0$. However, if electrostatic correlations dominate over other non-electrostatic ion-ion interactions, we expect $\alpha<0$ and $\chi<0$. Within the framework of the present model we obtain a bulk instability for this case. An extension of the present model to include stabilizing higher derivatives of the concentrations $\phi_a$ and $\phi_c$ in the free energy will likely constitute a meaningful model that also covers the case $\alpha<0$ and $\chi<0$.

\section*{Acknowledgments}
G.V. Bossa acknowledges a post-doctoral fellowship from Sao Paulo Research Foundation (FAPESP, Grant No. 2017/21772-2).


\begin{thebibliography}{48}%
\makeatletter
\providecommand \@ifxundefined [1]{%
 \@ifx{#1\undefined}
}%
\providecommand \@ifnum [1]{%
 \ifnum #1\expandafter \@firstoftwo
 \else \expandafter \@secondoftwo
 \fi
}%
\providecommand \@ifx [1]{%
 \ifx #1\expandafter \@firstoftwo
 \else \expandafter \@secondoftwo
 \fi
}%
\providecommand \natexlab [1]{#1}%
\providecommand \enquote  [1]{``#1''}%
\providecommand \bibnamefont  [1]{#1}%
\providecommand \bibfnamefont [1]{#1}%
\providecommand \citenamefont [1]{#1}%
\providecommand \href@noop [0]{\@secondoftwo}%
\providecommand \href [0]{\begingroup \@sanitize@url \@href}%
\providecommand \@href[1]{\@@startlink{#1}\@@href}%
\providecommand \@@href[1]{\endgroup#1\@@endlink}%
\providecommand \@sanitize@url [0]{\catcode `\\12\catcode `\$12\catcode
  `\&12\catcode `\#12\catcode `\^12\catcode `\_12\catcode `\%12\relax}%
\providecommand \@@startlink[1]{}%
\providecommand \@@endlink[0]{}%
\providecommand \url  [0]{\begingroup\@sanitize@url \@url }%
\providecommand \@url [1]{\endgroup\@href {#1}{\urlprefix }}%
\providecommand \urlprefix  [0]{URL }%
\providecommand \Eprint [0]{\href }%
\providecommand \doibase [0]{http://dx.doi.org/}%
\providecommand \selectlanguage [0]{\@gobble}%
\providecommand \bibinfo  [0]{\@secondoftwo}%
\providecommand \bibfield  [0]{\@secondoftwo}%
\providecommand \translation [1]{[#1]}%
\providecommand \BibitemOpen [0]{}%
\providecommand \bibitemStop [0]{}%
\providecommand \bibitemNoStop [0]{.\EOS\space}%
\providecommand \EOS [0]{\spacefactor3000\relax}%
\providecommand \BibitemShut  [1]{\csname bibitem#1\endcsname}%
\let\auto@bib@innerbib\@empty
\bibitem [{\citenamefont {Wishart}(2009)}]{wishart09}%
  \BibitemOpen
  \bibfield  {author} {\bibinfo {author} {\bibfnamefont {J.~F.}\ \bibnamefont
  {Wishart}},\ }\bibfield  {title} {\enquote {\bibinfo {title} {Energy
  applications of ionic liquids},}\ }\href@noop {} {\bibfield  {journal}
  {\bibinfo  {journal} {Energy Environ. Sci.}\ }\textbf {\bibinfo {volume}
  {2}},\ \bibinfo {pages} {956--961} (\bibinfo {year} {2009})}\BibitemShut
  {NoStop}%
\bibitem [{\citenamefont {Armand}\ \emph {et~al.}(2011)\citenamefont {Armand},
  \citenamefont {Endres}, \citenamefont {MacFarlane}, \citenamefont {Ohno},\
  and\ \citenamefont {Scrosati}}]{armand11}%
  \BibitemOpen
  \bibfield  {author} {\bibinfo {author} {\bibfnamefont {M.}~\bibnamefont
  {Armand}}, \bibinfo {author} {\bibfnamefont {F.}~\bibnamefont {Endres}},
  \bibinfo {author} {\bibfnamefont {D.~R.}\ \bibnamefont {MacFarlane}},
  \bibinfo {author} {\bibfnamefont {H.}~\bibnamefont {Ohno}}, \ and\ \bibinfo
  {author} {\bibfnamefont {B.}~\bibnamefont {Scrosati}},\ }\bibfield  {title}
  {\enquote {\bibinfo {title} {Ionic-liquid materials for the electrochemical
  challenges of the future},}\ }in\ \href@noop {} {\emph {\bibinfo {booktitle}
  {Materials For Sustainable Energy: A Collection of Peer-Reviewed Research and
  Review Articles from Nature Publishing Group}}}\ (\bibinfo  {publisher}
  {World Scientific},\ \bibinfo {year} {2011})\ pp.\ \bibinfo {pages}
  {129--137}\BibitemShut {NoStop}%
\bibitem [{\citenamefont {Watanabe}\ \emph {et~al.}(2017)\citenamefont
  {Watanabe}, \citenamefont {Thomas}, \citenamefont {Zhang}, \citenamefont
  {Ueno}, \citenamefont {Yasuda},\ and\ \citenamefont {Dokko}}]{watanabe17}%
  \BibitemOpen
  \bibfield  {author} {\bibinfo {author} {\bibfnamefont {M.}~\bibnamefont
  {Watanabe}}, \bibinfo {author} {\bibfnamefont {M.~L.}\ \bibnamefont
  {Thomas}}, \bibinfo {author} {\bibfnamefont {S.}~\bibnamefont {Zhang}},
  \bibinfo {author} {\bibfnamefont {K.}~\bibnamefont {Ueno}}, \bibinfo {author}
  {\bibfnamefont {T.}~\bibnamefont {Yasuda}}, \ and\ \bibinfo {author}
  {\bibfnamefont {K.}~\bibnamefont {Dokko}},\ }\bibfield  {title} {\enquote
  {\bibinfo {title} {Application of ionic liquids to energy storage and
  conversion materials and devices},}\ }\href@noop {} {\bibfield  {journal}
  {\bibinfo  {journal} {Chem. Rev.}\ }\textbf {\bibinfo {volume} {117}},\
  \bibinfo {pages} {7190--7239} (\bibinfo {year} {2017})}\BibitemShut {NoStop}%
\bibitem [{\citenamefont {Hayes}, \citenamefont {Warr},\ and\ \citenamefont
  {Atkin}(2015)}]{hayes15}%
  \BibitemOpen
  \bibfield  {author} {\bibinfo {author} {\bibfnamefont {R.}~\bibnamefont
  {Hayes}}, \bibinfo {author} {\bibfnamefont {G.~G.}\ \bibnamefont {Warr}}, \
  and\ \bibinfo {author} {\bibfnamefont {R.}~\bibnamefont {Atkin}},\ }\bibfield
   {title} {\enquote {\bibinfo {title} {Structure and nanostructure in ionic
  liquids},}\ }\href@noop {} {\bibfield  {journal} {\bibinfo  {journal}
  {Chemical reviews}\ }\textbf {\bibinfo {volume} {115}},\ \bibinfo {pages}
  {6357--6426} (\bibinfo {year} {2015})}\BibitemShut {NoStop}%
\bibitem [{\citenamefont {Balducci}\ \emph {et~al.}(2007)\citenamefont
  {Balducci}, \citenamefont {Dugas}, \citenamefont {Taberna}, \citenamefont
  {Simon}, \citenamefont {Plee}, \citenamefont {Mastragostino},\ and\
  \citenamefont {Passerini}}]{balducci07}%
  \BibitemOpen
  \bibfield  {author} {\bibinfo {author} {\bibfnamefont {A.}~\bibnamefont
  {Balducci}}, \bibinfo {author} {\bibfnamefont {R.}~\bibnamefont {Dugas}},
  \bibinfo {author} {\bibfnamefont {P.-L.}\ \bibnamefont {Taberna}}, \bibinfo
  {author} {\bibfnamefont {P.}~\bibnamefont {Simon}}, \bibinfo {author}
  {\bibfnamefont {D.}~\bibnamefont {Plee}}, \bibinfo {author} {\bibfnamefont
  {M.}~\bibnamefont {Mastragostino}}, \ and\ \bibinfo {author} {\bibfnamefont
  {S.}~\bibnamefont {Passerini}},\ }\bibfield  {title} {\enquote {\bibinfo
  {title} {High temperature carbon--carbon supercapacitor using ionic liquid as
  electrolyte},}\ }\href@noop {} {\bibfield  {journal} {\bibinfo  {journal} {J.
  Power Sources}\ }\textbf {\bibinfo {volume} {165}},\ \bibinfo {pages}
  {922--927} (\bibinfo {year} {2007})}\BibitemShut {NoStop}%
\bibitem [{\citenamefont {Armand}\ \emph {et~al.}(2009)\citenamefont {Armand},
  \citenamefont {Endres}, \citenamefont {MacFarlane}, \citenamefont {Ohno},\
  and\ \citenamefont {Scrosati}}]{armand09}%
  \BibitemOpen
  \bibfield  {author} {\bibinfo {author} {\bibfnamefont {M.}~\bibnamefont
  {Armand}}, \bibinfo {author} {\bibfnamefont {F.}~\bibnamefont {Endres}},
  \bibinfo {author} {\bibfnamefont {D.~R.}\ \bibnamefont {MacFarlane}},
  \bibinfo {author} {\bibfnamefont {H.}~\bibnamefont {Ohno}}, \ and\ \bibinfo
  {author} {\bibfnamefont {B.}~\bibnamefont {Scrosati}},\ }\bibfield  {title}
  {\enquote {\bibinfo {title} {Ionic-liquid materials for the electrochemical
  challenges of the future},}\ }\href@noop {} {\bibfield  {journal} {\bibinfo
  {journal} {Nat. Mater.}\ }\textbf {\bibinfo {volume} {8}},\ \bibinfo {pages}
  {621} (\bibinfo {year} {2009})}\BibitemShut {NoStop}%
\bibitem [{\citenamefont {MacFarlane}\ \emph {et~al.}(2014)\citenamefont
  {MacFarlane}, \citenamefont {Tachikawa}, \citenamefont {Forsyth},
  \citenamefont {Pringle}, \citenamefont {Howlett}, \citenamefont {Elliott},
  \citenamefont {Davis}, \citenamefont {Watanabe}, \citenamefont {Simon},\ and\
  \citenamefont {Angell}}]{macfarlane14}%
  \BibitemOpen
  \bibfield  {author} {\bibinfo {author} {\bibfnamefont {D.~R.}\ \bibnamefont
  {MacFarlane}}, \bibinfo {author} {\bibfnamefont {N.}~\bibnamefont
  {Tachikawa}}, \bibinfo {author} {\bibfnamefont {M.}~\bibnamefont {Forsyth}},
  \bibinfo {author} {\bibfnamefont {J.~M.}\ \bibnamefont {Pringle}}, \bibinfo
  {author} {\bibfnamefont {P.~C.}\ \bibnamefont {Howlett}}, \bibinfo {author}
  {\bibfnamefont {G.~D.}\ \bibnamefont {Elliott}}, \bibinfo {author}
  {\bibfnamefont {J.~H.}\ \bibnamefont {Davis}}, \bibinfo {author}
  {\bibfnamefont {M.}~\bibnamefont {Watanabe}}, \bibinfo {author}
  {\bibfnamefont {P.}~\bibnamefont {Simon}}, \ and\ \bibinfo {author}
  {\bibfnamefont {C.~A.}\ \bibnamefont {Angell}},\ }\bibfield  {title}
  {\enquote {\bibinfo {title} {Energy applications of ionic liquids},}\
  }\href@noop {} {\bibfield  {journal} {\bibinfo  {journal} {Energy Environ.
  Sci.}\ }\textbf {\bibinfo {volume} {7}},\ \bibinfo {pages} {232--250}
  (\bibinfo {year} {2014})}\BibitemShut {NoStop}%
\bibitem [{\citenamefont {Salanne}(2017)}]{salanne17}%
  \BibitemOpen
  \bibfield  {author} {\bibinfo {author} {\bibfnamefont {M.}~\bibnamefont
  {Salanne}},\ }\bibfield  {title} {\enquote {\bibinfo {title} {Ionic liquids
  for supercapacitor applications},}\ }\href@noop {} {\bibfield  {journal}
  {\bibinfo  {journal} {Top. Curr. Chem.}\ }\textbf {\bibinfo {volume} {375}},\
  \bibinfo {pages} {63} (\bibinfo {year} {2017})}\BibitemShut {NoStop}%
\bibitem [{\citenamefont {Fedorov}\ and\ \citenamefont
  {Kornyshev}(2014)}]{fedorov14}%
  \BibitemOpen
  \bibfield  {author} {\bibinfo {author} {\bibfnamefont {M.~V.}\ \bibnamefont
  {Fedorov}}\ and\ \bibinfo {author} {\bibfnamefont {A.~A.}\ \bibnamefont
  {Kornyshev}},\ }\bibfield  {title} {\enquote {\bibinfo {title} {Ionic liquids
  at electrified interfaces},}\ }\href@noop {} {\bibfield  {journal} {\bibinfo
  {journal} {Chem. Rev.}\ }\textbf {\bibinfo {volume} {114}},\ \bibinfo {pages}
  {2978--3036} (\bibinfo {year} {2014})}\BibitemShut {NoStop}%
\bibitem [{\citenamefont {Bikerman}(1942)}]{bikerman42}%
  \BibitemOpen
  \bibfield  {author} {\bibinfo {author} {\bibfnamefont {J.}~\bibnamefont
  {Bikerman}},\ }\bibfield  {title} {\enquote {\bibinfo {title} {Structure and
  capacity of electrical double layer},}\ }\href@noop {} {\bibfield  {journal}
  {\bibinfo  {journal} {Philos. Mag. (1798-1977)}\ }\textbf {\bibinfo {volume}
  {33}},\ \bibinfo {pages} {384--397} (\bibinfo {year} {1942})}\BibitemShut
  {NoStop}%
\bibitem [{\citenamefont {Kornyshev}\ \emph {et~al.}(2007)\citenamefont
  {Kornyshev} \emph {et~al.}}]{kornyshev07}%
  \BibitemOpen
  \bibfield  {author} {\bibinfo {author} {\bibfnamefont {A.}~\bibnamefont
  {Kornyshev}} \emph {et~al.},\ }\bibfield  {title} {\enquote {\bibinfo {title}
  {Double-layer in ionic liquids: Paradigm change?}}\ }\href@noop {} {\bibfield
   {journal} {\bibinfo  {journal} {J. Phys. Chem. B}\ }\textbf {\bibinfo
  {volume} {111}},\ \bibinfo {pages} {5545--5557} (\bibinfo {year}
  {2007})}\BibitemShut {NoStop}%
\bibitem [{\citenamefont {Popovi{\'c}}\ and\ \citenamefont
  {{\v{S}}iber}(2013)}]{popovic13}%
  \BibitemOpen
  \bibfield  {author} {\bibinfo {author} {\bibfnamefont {M.}~\bibnamefont
  {Popovi{\'c}}}\ and\ \bibinfo {author} {\bibfnamefont {A.}~\bibnamefont
  {{\v{S}}iber}},\ }\bibfield  {title} {\enquote {\bibinfo {title} {Lattice-gas
  {Poisson-Boltzmann} approach for sterically asymmetric electrolytes},}\
  }\href@noop {} {\bibfield  {journal} {\bibinfo  {journal} {Phys. Rev. E}\
  }\textbf {\bibinfo {volume} {88}},\ \bibinfo {pages} {022302} (\bibinfo
  {year} {2013})}\BibitemShut {NoStop}%
\bibitem [{\citenamefont {Han}, \citenamefont {Huang},\ and\ \citenamefont
  {Yan}(2014)}]{han14}%
  \BibitemOpen
  \bibfield  {author} {\bibinfo {author} {\bibfnamefont {Y.}~\bibnamefont
  {Han}}, \bibinfo {author} {\bibfnamefont {S.}~\bibnamefont {Huang}}, \ and\
  \bibinfo {author} {\bibfnamefont {T.}~\bibnamefont {Yan}},\ }\bibfield
  {title} {\enquote {\bibinfo {title} {A mean-field theory on the differential
  capacitance of asymmetric ionic liquid electrolytes},}\ }\href@noop {}
  {\bibfield  {journal} {\bibinfo  {journal} {J. Phys.: Condens. Matter}\
  }\textbf {\bibinfo {volume} {26}},\ \bibinfo {pages} {284103} (\bibinfo
  {year} {2014})}\BibitemShut {NoStop}%
\bibitem [{\citenamefont {Girotto}\ \emph {et~al.}(2017)\citenamefont
  {Girotto}, \citenamefont {Colla}, \citenamefont {dos Santos},\ and\
  \citenamefont {Levin}}]{girotto17}%
  \BibitemOpen
  \bibfield  {author} {\bibinfo {author} {\bibfnamefont {M.}~\bibnamefont
  {Girotto}}, \bibinfo {author} {\bibfnamefont {T.}~\bibnamefont {Colla}},
  \bibinfo {author} {\bibfnamefont {A.~P.}\ \bibnamefont {dos Santos}}, \ and\
  \bibinfo {author} {\bibfnamefont {Y.}~\bibnamefont {Levin}},\ }\bibfield
  {title} {\enquote {\bibinfo {title} {Lattice model of an ionic liquid at an
  electrified interface},}\ }\href@noop {} {\bibfield  {journal} {\bibinfo
  {journal} {J. Phys. Chem. B}\ }\textbf {\bibinfo {volume} {121}},\ \bibinfo
  {pages} {6408--6415} (\bibinfo {year} {2017})}\BibitemShut {NoStop}%
\bibitem [{\citenamefont {Lockett}\ \emph {et~al.}(2010)\citenamefont
  {Lockett}, \citenamefont {Horne}, \citenamefont {Sedev}, \citenamefont
  {Rodopoulos},\ and\ \citenamefont {Ralston}}]{lockett10}%
  \BibitemOpen
  \bibfield  {author} {\bibinfo {author} {\bibfnamefont {V.}~\bibnamefont
  {Lockett}}, \bibinfo {author} {\bibfnamefont {M.}~\bibnamefont {Horne}},
  \bibinfo {author} {\bibfnamefont {R.}~\bibnamefont {Sedev}}, \bibinfo
  {author} {\bibfnamefont {T.}~\bibnamefont {Rodopoulos}}, \ and\ \bibinfo
  {author} {\bibfnamefont {J.}~\bibnamefont {Ralston}},\ }\bibfield  {title}
  {\enquote {\bibinfo {title} {Differential capacitance of the double layer at
  the electrode/ionic liquids interface},}\ }\href@noop {} {\bibfield
  {journal} {\bibinfo  {journal} {Phys. Chem. Chem. Phys.}\ }\textbf {\bibinfo
  {volume} {12}},\ \bibinfo {pages} {12499--12512} (\bibinfo {year}
  {2010})}\BibitemShut {NoStop}%
\bibitem [{\citenamefont {Lamperski}, \citenamefont {Outhwaite},\ and\
  \citenamefont {Bhuiyan}(2009)}]{lamperski09}%
  \BibitemOpen
  \bibfield  {author} {\bibinfo {author} {\bibfnamefont {S.}~\bibnamefont
  {Lamperski}}, \bibinfo {author} {\bibfnamefont {C.~W.}\ \bibnamefont
  {Outhwaite}}, \ and\ \bibinfo {author} {\bibfnamefont {L.~B.}\ \bibnamefont
  {Bhuiyan}},\ }\bibfield  {title} {\enquote {\bibinfo {title} {The electric
  double-layer differential capacitance at and near zero surface charge for a
  restricted primitive model electrolyte},}\ }\href@noop {} {\bibfield
  {journal} {\bibinfo  {journal} {J. Phys. Chem. B}\ }\textbf {\bibinfo
  {volume} {113}},\ \bibinfo {pages} {8925--8929} (\bibinfo {year}
  {2009})}\BibitemShut {NoStop}%
\bibitem [{\citenamefont {Caetano}\ \emph {et~al.}(2016)\citenamefont
  {Caetano}, \citenamefont {Bossa}, \citenamefont {de~Oliveira}, \citenamefont
  {Brown}, \citenamefont {de~Carvalho},\ and\ \citenamefont {May}}]{caetano16}%
  \BibitemOpen
  \bibfield  {author} {\bibinfo {author} {\bibfnamefont {D.~L.}\ \bibnamefont
  {Caetano}}, \bibinfo {author} {\bibfnamefont {G.~V.}\ \bibnamefont {Bossa}},
  \bibinfo {author} {\bibfnamefont {V.~M.}\ \bibnamefont {de~Oliveira}},
  \bibinfo {author} {\bibfnamefont {M.~A.}\ \bibnamefont {Brown}}, \bibinfo
  {author} {\bibfnamefont {S.~J.}\ \bibnamefont {de~Carvalho}}, \ and\ \bibinfo
  {author} {\bibfnamefont {S.}~\bibnamefont {May}},\ }\bibfield  {title}
  {\enquote {\bibinfo {title} {Role of ion hydration for the differential
  capacitance of an electric double layer},}\ }\href@noop {} {\bibfield
  {journal} {\bibinfo  {journal} {Phys. Chem. Chem. Phys.}\ }\textbf {\bibinfo
  {volume} {18}},\ \bibinfo {pages} {27796--27807} (\bibinfo {year}
  {2016})}\BibitemShut {NoStop}%
\bibitem [{\citenamefont {Lue}, \citenamefont {Zoeller},\ and\ \citenamefont
  {Blankschtein}(1999)}]{lue99}%
  \BibitemOpen
  \bibfield  {author} {\bibinfo {author} {\bibfnamefont {L.}~\bibnamefont
  {Lue}}, \bibinfo {author} {\bibfnamefont {N.}~\bibnamefont {Zoeller}}, \ and\
  \bibinfo {author} {\bibfnamefont {D.}~\bibnamefont {Blankschtein}},\
  }\bibfield  {title} {\enquote {\bibinfo {title} {Incorporation of
  nonelectrostatic interactions in the {Poisson-Boltzmann} equation},}\
  }\href@noop {} {\bibfield  {journal} {\bibinfo  {journal} {Langmuir}\
  }\textbf {\bibinfo {volume} {15}},\ \bibinfo {pages} {3726--3730} (\bibinfo
  {year} {1999})}\BibitemShut {NoStop}%
\bibitem [{\citenamefont {Biesheuvel}\ and\ \citenamefont
  {Van~Soestbergen}(2007)}]{biesheuvel07}%
  \BibitemOpen
  \bibfield  {author} {\bibinfo {author} {\bibfnamefont {P.}~\bibnamefont
  {Biesheuvel}}\ and\ \bibinfo {author} {\bibfnamefont {M.}~\bibnamefont
  {Van~Soestbergen}},\ }\bibfield  {title} {\enquote {\bibinfo {title}
  {Counterion volume effects in mixed electrical double layers},}\ }\href@noop
  {} {\bibfield  {journal} {\bibinfo  {journal} {J. Colloid Interface Sci.}\
  }\textbf {\bibinfo {volume} {316}},\ \bibinfo {pages} {490--499} (\bibinfo
  {year} {2007})}\BibitemShut {NoStop}%
\bibitem [{\citenamefont {Gavish}\ and\ \citenamefont
  {Promislow}(2016)}]{gavish16}%
  \BibitemOpen
  \bibfield  {author} {\bibinfo {author} {\bibfnamefont {N.}~\bibnamefont
  {Gavish}}\ and\ \bibinfo {author} {\bibfnamefont {K.}~\bibnamefont
  {Promislow}},\ }\bibfield  {title} {\enquote {\bibinfo {title} {On the
  structure of generalized {Poisson-Boltzmann} equations},}\ }\href@noop {}
  {\bibfield  {journal} {\bibinfo  {journal} {Eur. J. Appl. Math.}\ }\textbf
  {\bibinfo {volume} {27}},\ \bibinfo {pages} {667--685} (\bibinfo {year}
  {2016})}\BibitemShut {NoStop}%
\bibitem [{\citenamefont {Maggs}\ and\ \citenamefont
  {Podgornik}(2016)}]{maggs16}%
  \BibitemOpen
  \bibfield  {author} {\bibinfo {author} {\bibfnamefont {A.}~\bibnamefont
  {Maggs}}\ and\ \bibinfo {author} {\bibfnamefont {R.}~\bibnamefont
  {Podgornik}},\ }\bibfield  {title} {\enquote {\bibinfo {title} {General
  theory of asymmetric steric interactions in electrostatic double layers},}\
  }\href@noop {} {\bibfield  {journal} {\bibinfo  {journal} {Soft Matter}\
  }\textbf {\bibinfo {volume} {12}},\ \bibinfo {pages} {1219--1229} (\bibinfo
  {year} {2016})}\BibitemShut {NoStop}%
\bibitem [{\citenamefont {Bazant}, \citenamefont {Storey},\ and\ \citenamefont
  {Kornyshev}(2011)}]{bazant11}%
  \BibitemOpen
  \bibfield  {author} {\bibinfo {author} {\bibfnamefont {M.~Z.}\ \bibnamefont
  {Bazant}}, \bibinfo {author} {\bibfnamefont {B.~D.}\ \bibnamefont {Storey}},
  \ and\ \bibinfo {author} {\bibfnamefont {A.~A.}\ \bibnamefont {Kornyshev}},\
  }\bibfield  {title} {\enquote {\bibinfo {title} {Double layer in ionic
  liquids: Overscreening versus crowding},}\ }\href@noop {} {\bibfield
  {journal} {\bibinfo  {journal} {Phys. Rev. Lett.}\ }\textbf {\bibinfo
  {volume} {106}},\ \bibinfo {pages} {046102} (\bibinfo {year}
  {2011})}\BibitemShut {NoStop}%
\bibitem [{\citenamefont {Aoki}(2012)}]{aoki12}%
  \BibitemOpen
  \bibfield  {author} {\bibinfo {author} {\bibfnamefont {K.}~\bibnamefont
  {Aoki}},\ }\bibfield  {title} {\enquote {\bibinfo {title} {Ion-cell model for
  electric double layers composed of rigid ions},}\ }\href@noop {} {\bibfield
  {journal} {\bibinfo  {journal} {Electrochim. Acta}\ }\textbf {\bibinfo
  {volume} {67}},\ \bibinfo {pages} {216--223} (\bibinfo {year}
  {2012})}\BibitemShut {NoStop}%
\bibitem [{\citenamefont {Goodwin}, \citenamefont {Feng},\ and\ \citenamefont
  {Kornyshev}(2017)}]{goodwin17}%
  \BibitemOpen
  \bibfield  {author} {\bibinfo {author} {\bibfnamefont {Z.~A.}\ \bibnamefont
  {Goodwin}}, \bibinfo {author} {\bibfnamefont {G.}~\bibnamefont {Feng}}, \
  and\ \bibinfo {author} {\bibfnamefont {A.~A.}\ \bibnamefont {Kornyshev}},\
  }\bibfield  {title} {\enquote {\bibinfo {title} {Mean-field theory of
  electrical double layer in ionic liquids with account of short-range
  correlations},}\ }\href@noop {} {\bibfield  {journal} {\bibinfo  {journal}
  {Electrochim. Acta}\ }\textbf {\bibinfo {volume} {225}},\ \bibinfo {pages}
  {190--197} (\bibinfo {year} {2017})}\BibitemShut {NoStop}%
\bibitem [{\citenamefont {Gavish}, \citenamefont {Elad},\ and\ \citenamefont
  {Yochelis}(2017)}]{gavish17}%
  \BibitemOpen
  \bibfield  {author} {\bibinfo {author} {\bibfnamefont {N.}~\bibnamefont
  {Gavish}}, \bibinfo {author} {\bibfnamefont {D.}~\bibnamefont {Elad}}, \ and\
  \bibinfo {author} {\bibfnamefont {A.}~\bibnamefont {Yochelis}},\ }\bibfield
  {title} {\enquote {\bibinfo {title} {From solvent-free to dilute
  electrolytes: essential components for a continuum theory},}\ }\href@noop {}
  {\bibfield  {journal} {\bibinfo  {journal} {J. Phys. Chem. Lett.}\ }\textbf
  {\bibinfo {volume} {9}},\ \bibinfo {pages} {36--42} (\bibinfo {year}
  {2017})}\BibitemShut {NoStop}%
\bibitem [{\citenamefont {Bokun}\ \emph {et~al.}(2018)\citenamefont {Bokun},
  \citenamefont {di~Caprio}, \citenamefont {Holovko},\ and\ \citenamefont
  {Vikhrenko}}]{bokun18}%
  \BibitemOpen
  \bibfield  {author} {\bibinfo {author} {\bibfnamefont {G.}~\bibnamefont
  {Bokun}}, \bibinfo {author} {\bibfnamefont {D.}~\bibnamefont {di~Caprio}},
  \bibinfo {author} {\bibfnamefont {M.}~\bibnamefont {Holovko}}, \ and\
  \bibinfo {author} {\bibfnamefont {V.}~\bibnamefont {Vikhrenko}},\ }\bibfield
  {title} {\enquote {\bibinfo {title} {The system of mobile ions in lattice
  models: Screening effects, thermodynamic and electrophysical properties},}\
  }\href@noop {} {\bibfield  {journal} {\bibinfo  {journal} {J. Mol. Liq.}\ }\textbf {\bibinfo {volume}
  {270}},\ \bibinfo {pages} {183--190} 
  (\bibinfo {year} {2018})}\BibitemShut {NoStop}%
\bibitem [{\citenamefont {Yin}\ \emph {et~al.}(2018)\citenamefont {Yin},
  \citenamefont {Huang}, \citenamefont {Chen},\ and\ \citenamefont
  {Yan}}]{yin18}%
  \BibitemOpen
  \bibfield  {author} {\bibinfo {author} {\bibfnamefont {L.}~\bibnamefont
  {Yin}}, \bibinfo {author} {\bibfnamefont {Y.}~\bibnamefont {Huang}}, \bibinfo
  {author} {\bibfnamefont {H.}~\bibnamefont {Chen}}, \ and\ \bibinfo {author}
  {\bibfnamefont {T.}~\bibnamefont {Yan}},\ }\bibfield  {title} {\enquote
  {\bibinfo {title} {A mean-field theory on the differential capacitance of
  asymmetric ionic liquid electrolytes. {II.} accounts of ionic
  interactions},}\ }\href@noop {} {\bibfield  {journal} {\bibinfo  {journal}
  {Phys. Chem. Chem. Phys.}\ }\textbf {\bibinfo {volume}
  {20}},\ \bibinfo {pages} {17606-17614} (\bibinfo {year} {2018})}\BibitemShut {NoStop}%
\bibitem [{\citenamefont {Rotenberg}, \citenamefont {Bernard},\ and\
  \citenamefont {Hansen}(2018)}]{rotenberg18}%
  \BibitemOpen
  \bibfield  {author} {\bibinfo {author} {\bibfnamefont {B.}~\bibnamefont
  {Rotenberg}}, \bibinfo {author} {\bibfnamefont {O.}~\bibnamefont {Bernard}},
  \ and\ \bibinfo {author} {\bibfnamefont {J.-P.}\ \bibnamefont {Hansen}},\
  }\bibfield  {title} {\enquote {\bibinfo {title} {Underscreening in ionic
  liquids: a first principles analysis},}\ }\href@noop {} {\bibfield  {journal}
  {\bibinfo  {journal} {J. Phys.: Condens. Matter}\ }\textbf {\bibinfo {volume}
  {30}},\ \bibinfo {pages} {054005} (\bibinfo {year} {2018})}\BibitemShut
  {NoStop}%
\bibitem [{\citenamefont {Perkin}(2012)}]{perkin12}%
  \BibitemOpen
  \bibfield  {author} {\bibinfo {author} {\bibfnamefont {S.}~\bibnamefont
  {Perkin}},\ }\bibfield  {title} {\enquote {\bibinfo {title} {Ionic liquids in
  confined geometries},}\ }\href@noop {} {\bibfield  {journal} {\bibinfo
  {journal} {Phys. Chem. Chem. Phys.}\ }\textbf {\bibinfo {volume} {14}},\
  \bibinfo {pages} {5052--5062} (\bibinfo {year} {2012})}\BibitemShut {NoStop}%
\bibitem [{\citenamefont {Gebbie}\ \emph {et~al.}(2013)\citenamefont {Gebbie},
  \citenamefont {Valtiner}, \citenamefont {Banquy}, \citenamefont {Fox},
  \citenamefont {Henderson},\ and\ \citenamefont {Israelachvili}}]{gebbie13}%
  \BibitemOpen
  \bibfield  {author} {\bibinfo {author} {\bibfnamefont {M.~A.}\ \bibnamefont
  {Gebbie}}, \bibinfo {author} {\bibfnamefont {M.}~\bibnamefont {Valtiner}},
  \bibinfo {author} {\bibfnamefont {X.}~\bibnamefont {Banquy}}, \bibinfo
  {author} {\bibfnamefont {E.~T.}\ \bibnamefont {Fox}}, \bibinfo {author}
  {\bibfnamefont {W.~A.}\ \bibnamefont {Henderson}}, \ and\ \bibinfo {author}
  {\bibfnamefont {J.~N.}\ \bibnamefont {Israelachvili}},\ }\bibfield  {title}
  {\enquote {\bibinfo {title} {Ionic liquids behave as dilute electrolyte
  solutions},}\ }\href@noop {} {\bibfield  {journal} {\bibinfo  {journal}
  {Proc. Natl. Acad. Sci. U. S. A.}\ }\textbf {\bibinfo {volume} {110}},\
  \bibinfo {pages} {9674--9679} (\bibinfo {year} {2013})}\BibitemShut {NoStop}%
\bibitem [{\citenamefont {Gebbie}\ \emph {et~al.}(2017)\citenamefont {Gebbie},
  \citenamefont {Smith}, \citenamefont {Dobbs}, \citenamefont {Warr},
  \citenamefont {Banquy}, \citenamefont {Valtiner}, \citenamefont {Rutland},
  \citenamefont {Israelachvili}, \citenamefont {Perkin}, \citenamefont {Atkin}
  \emph {et~al.}}]{gebbie17}%
  \BibitemOpen
  \bibfield  {author} {\bibinfo {author} {\bibfnamefont {M.~A.}\ \bibnamefont
  {Gebbie}}, \bibinfo {author} {\bibfnamefont {A.~M.}\ \bibnamefont {Smith}},
  \bibinfo {author} {\bibfnamefont {H.~A.}\ \bibnamefont {Dobbs}}, \bibinfo
  {author} {\bibfnamefont {G.~G.}\ \bibnamefont {Warr}}, \bibinfo {author}
  {\bibfnamefont {X.}~\bibnamefont {Banquy}}, \bibinfo {author} {\bibfnamefont
  {M.}~\bibnamefont {Valtiner}}, \bibinfo {author} {\bibfnamefont {M.~W.}\
  \bibnamefont {Rutland}}, \bibinfo {author} {\bibfnamefont {J.~N.}\
  \bibnamefont {Israelachvili}}, \bibinfo {author} {\bibfnamefont
  {S.}~\bibnamefont {Perkin}}, \bibinfo {author} {\bibfnamefont
  {R.}~\bibnamefont {Atkin}},  \emph {et~al.},\ }\bibfield  {title} {\enquote
  {\bibinfo {title} {Long range electrostatic forces in ionic liquids},}\
  }\href@noop {} {\bibfield  {journal} {\bibinfo  {journal} {Chem. Commun.}\
  }\textbf {\bibinfo {volume} {53}},\ \bibinfo {pages} {1214--1224} (\bibinfo
  {year} {2017})}\BibitemShut {NoStop}%
\bibitem [{\citenamefont {Perez-Martinez}\ \emph {et~al.}(2017)\citenamefont
  {Perez-Martinez}, \citenamefont {Smith}, \citenamefont {Perkin} \emph
  {et~al.}}]{perez17}%
  \BibitemOpen
  \bibfield  {author} {\bibinfo {author} {\bibfnamefont {C.~S.}\ \bibnamefont
  {Perez-Martinez}}, \bibinfo {author} {\bibfnamefont {A.~M.}\ \bibnamefont
  {Smith}}, \bibinfo {author} {\bibfnamefont {S.}~\bibnamefont {Perkin}},
  \emph {et~al.},\ }\bibfield  {title} {\enquote {\bibinfo {title}
  {Underscreening in concentrated electrolytes},}\ }\href@noop {} {\bibfield
  {journal} {\bibinfo  {journal} {Faraday Discuss.}\ }\textbf {\bibinfo
  {volume} {199}},\ \bibinfo {pages} {239--259} (\bibinfo {year}
  {2017})}\BibitemShut {NoStop}%
\bibitem [{\citenamefont {Goodwin}\ and\ \citenamefont
  {Kornyshev}(2017)}]{kornyshev17}%
  \BibitemOpen
  \bibfield  {author} {\bibinfo {author} {\bibfnamefont {Z.~A.}\ \bibnamefont
  {Goodwin}}\ and\ \bibinfo {author} {\bibfnamefont {A.~A.}\ \bibnamefont
  {Kornyshev}},\ }\bibfield  {title} {\enquote {\bibinfo {title}
  {Underscreening, overscreening and double-layer capacitance},}\ }\href@noop
  {} {\bibfield  {journal} {\bibinfo  {journal} {Electrochem. Commun.}\
  }\textbf {\bibinfo {volume} {82}},\ \bibinfo {pages} {129--133} (\bibinfo
  {year} {2017})}\BibitemShut {NoStop}%
\bibitem [{\citenamefont {Blossey}, \citenamefont {Maggs},\ and\ \citenamefont
  {Podgornik}(2017)}]{blossey17}%
  \BibitemOpen
  \bibfield  {author} {\bibinfo {author} {\bibfnamefont {R.}~\bibnamefont
  {Blossey}}, \bibinfo {author} {\bibfnamefont {A.}~\bibnamefont {Maggs}}, \
  and\ \bibinfo {author} {\bibfnamefont {R.}~\bibnamefont {Podgornik}},\
  }\bibfield  {title} {\enquote {\bibinfo {title} {Structural interactions in
  ionic liquids linked to higher-order {Poisson-Boltzmann} equations},}\
  }\href@noop {} {\bibfield  {journal} {\bibinfo  {journal} {Phys. Rev. E}\
  }\textbf {\bibinfo {volume} {95}},\ \bibinfo {pages} {060602} (\bibinfo
  {year} {2017})}\BibitemShut {NoStop}%
\bibitem [{\citenamefont {Davis}(1996)}]{davis96}%
  \BibitemOpen
  \bibfield  {author} {\bibinfo {author} {\bibfnamefont {H.~T.}\ \bibnamefont
  {Davis}},\ }\href@noop {} {\emph {\bibinfo {title} {Statistical mechanics of
  phases, interfaces, and thin films}}}\ (\bibinfo  {publisher} {VCH New
  York},\ \bibinfo {year} {1996})\BibitemShut {NoStop}%
\bibitem [{\citenamefont {Doi}(2013)}]{doi13}%
  \BibitemOpen
  \bibfield  {author} {\bibinfo {author} {\bibfnamefont {M.}~\bibnamefont
  {Doi}},\ }\href@noop {} {\emph {\bibinfo {title} {Soft Matter Physics}}}\
  (\bibinfo  {publisher} {Oxford University Press},\ \bibinfo {year}
  {2013})\BibitemShut {NoStop}%
\bibitem [{\citenamefont {Partenskii}\ and\ \citenamefont
  {Jordan}(2008)}]{partenskii08}%
  \BibitemOpen
  \bibfield  {author} {\bibinfo {author} {\bibfnamefont {M.~B.}\ \bibnamefont
  {Partenskii}}\ and\ \bibinfo {author} {\bibfnamefont {P.~C.}\ \bibnamefont
  {Jordan}},\ }\bibfield  {title} {\enquote {\bibinfo {title} {Limitations and
  strengths of uniformly charged double-layer theory: Physical significance of
  capacitance anomalies},}\ }\href@noop {} {\bibfield  {journal} {\bibinfo
  {journal} {Phys. Rev, E}\ }\textbf {\bibinfo {volume} {77}},\ \bibinfo
  {pages} {061117} (\bibinfo {year} {2008})}\BibitemShut {NoStop}%
\bibitem [{\citenamefont {Partenskii}\ and\ \citenamefont
  {Jordan}(2011)}]{partenskii11}%
  \BibitemOpen
  \bibfield  {author} {\bibinfo {author} {\bibfnamefont {M.}~\bibnamefont
  {Partenskii}}\ and\ \bibinfo {author} {\bibfnamefont {P.}~\bibnamefont
  {Jordan}},\ }\bibfield  {title} {\enquote {\bibinfo {title} {Relaxing gap
  capacitor models of electrified interfaces},}\ }\href@noop {} {\bibfield
  {journal} {\bibinfo  {journal} {Am. J. Phys.}\ }\textbf {\bibinfo {volume}
  {79}},\ \bibinfo {pages} {103--110} (\bibinfo {year} {2011})}\BibitemShut
  {NoStop}%
\bibitem [{\citenamefont {Kondrat}\ and\ \citenamefont
  {Kornyshev}(2016)}]{kondrat16}%
  \BibitemOpen
  \bibfield  {author} {\bibinfo {author} {\bibfnamefont {S.}~\bibnamefont
  {Kondrat}}\ and\ \bibinfo {author} {\bibfnamefont {A.~A.}\ \bibnamefont
  {Kornyshev}},\ }\bibfield  {title} {\enquote {\bibinfo {title} {Pressing a
  spring: what does it take to maximize the energy storage in nanoporous
  supercapacitors?}}\ }\href@noop {} {\bibfield  {journal} {\bibinfo  {journal}
  {Nanoscale Horiz.}\ }\textbf {\bibinfo {volume} {1}},\ \bibinfo {pages}
  {45--52} (\bibinfo {year} {2016})}\BibitemShut {NoStop}%
\bibitem [{\citenamefont {Lauw}\ \emph {et~al.}(2012)\citenamefont {Lauw},
  \citenamefont {Horne}, \citenamefont {Rodopoulos}, \citenamefont {Lockett},
  \citenamefont {Akgun}, \citenamefont {Hamilton},\ and\ \citenamefont
  {Nelson}}]{lauw12}%
  \BibitemOpen
  \bibfield  {author} {\bibinfo {author} {\bibfnamefont {Y.}~\bibnamefont
  {Lauw}}, \bibinfo {author} {\bibfnamefont {M.~D.}\ \bibnamefont {Horne}},
  \bibinfo {author} {\bibfnamefont {T.}~\bibnamefont {Rodopoulos}}, \bibinfo
  {author} {\bibfnamefont {V.}~\bibnamefont {Lockett}}, \bibinfo {author}
  {\bibfnamefont {B.}~\bibnamefont {Akgun}}, \bibinfo {author} {\bibfnamefont
  {W.~A.}\ \bibnamefont {Hamilton}}, \ and\ \bibinfo {author} {\bibfnamefont
  {A.~R.}\ \bibnamefont {Nelson}},\ }\bibfield  {title} {\enquote {\bibinfo
  {title} {Structure of [c4mpyr][ntf2] room-temperature ionic liquid at charged
  gold interfaces},}\ }\href@noop {} {\bibfield  {journal} {\bibinfo  {journal}
  {Langmuir}\ }\textbf {\bibinfo {volume} {28}},\ \bibinfo {pages} {7374--7381}
  (\bibinfo {year} {2012})}\BibitemShut {NoStop}%
\bibitem [{\citenamefont {Jitvisate}\ and\ \citenamefont
  {Seddon}(2017)}]{seddon18}%
  \BibitemOpen
  \bibfield  {author} {\bibinfo {author} {\bibfnamefont {M.}~\bibnamefont
  {Jitvisate}}\ and\ \bibinfo {author} {\bibfnamefont {J.~R.}\ \bibnamefont
  {Seddon}},\ }\bibfield  {title} {\enquote {\bibinfo {title} {Direct
  measurement of the differential capacitance of solvent-free and dilute ionic
  liquids},}\ }\href@noop {} {\bibfield  {journal} {\bibinfo  {journal} {J.
  Phys. Chem. Lett.}\ }\textbf {\bibinfo {volume} {9}},\ \bibinfo {pages}
  {126--131} (\bibinfo {year} {2017})}\BibitemShut {NoStop}%
\bibitem [{\citenamefont {Kilic}, \citenamefont {Bazant},\ and\ \citenamefont
  {Ajdari}(2007)}]{kilic07}%
  \BibitemOpen
  \bibfield  {author} {\bibinfo {author} {\bibfnamefont {M.~S.}\ \bibnamefont
  {Kilic}}, \bibinfo {author} {\bibfnamefont {M.~Z.}\ \bibnamefont {Bazant}}, \
  and\ \bibinfo {author} {\bibfnamefont {A.}~\bibnamefont {Ajdari}},\
  }\bibfield  {title} {\enquote {\bibinfo {title} {Steric effects in the
  dynamics of electrolytes at large applied voltages. {I.}~double-layer
  charging},}\ }\href@noop {} {\bibfield  {journal} {\bibinfo  {journal} {Phys.
  Rev. E}\ }\textbf {\bibinfo {volume} {75}},\ \bibinfo {pages} {021502}
  (\bibinfo {year} {2007})}\BibitemShut {NoStop}%
\bibitem [{\citenamefont {Torrie}(1992)}]{torrie92}%
  \BibitemOpen
  \bibfield  {author} {\bibinfo {author} {\bibfnamefont {G.}~\bibnamefont
  {Torrie}},\ }\bibfield  {title} {\enquote {\bibinfo {title} {Negative
  differential capacities in electrical double layers},}\ }\href@noop {}
  {\bibfield  {journal} {\bibinfo  {journal} {J. Chem. Phys.}\ }\textbf
  {\bibinfo {volume} {96}},\ \bibinfo {pages} {3772--3774} (\bibinfo {year}
  {1992})}\BibitemShut {NoStop}%
\bibitem [{\citenamefont {Boda}\ \emph {et~al.}(2002)\citenamefont {Boda},
  \citenamefont {Fawcett}, \citenamefont {Henderson},\ and\ \citenamefont
  {Soko{\l}owski}}]{boda02}%
  \BibitemOpen
  \bibfield  {author} {\bibinfo {author} {\bibfnamefont {D.}~\bibnamefont
  {Boda}}, \bibinfo {author} {\bibfnamefont {W.~R.}\ \bibnamefont {Fawcett}},
  \bibinfo {author} {\bibfnamefont {D.}~\bibnamefont {Henderson}}, \ and\
  \bibinfo {author} {\bibfnamefont {S.}~\bibnamefont {Soko{\l}owski}},\
  }\bibfield  {title} {\enquote {\bibinfo {title} {{Monte Carlo}, density
  functional theory, and {Poisson--Boltzmann} theory study of the structure of
  an electrolyte near an electrode},}\ }\href@noop {} {\bibfield  {journal}
  {\bibinfo  {journal} {J. Chem. Phys.}\ }\textbf {\bibinfo {volume} {116}},\
  \bibinfo {pages} {7170--7176} (\bibinfo {year} {2002})}\BibitemShut {NoStop}%
\bibitem [{\citenamefont {Boda}\ \emph {et~al.}(2004)\citenamefont {Boda},
  \citenamefont {Henderson}, \citenamefont {Plaschko},\ and\ \citenamefont
  {Ronald~Fawcett}}]{boda04}%
  \BibitemOpen
  \bibfield  {author} {\bibinfo {author} {\bibfnamefont {D.}~\bibnamefont
  {Boda}}, \bibinfo {author} {\bibfnamefont {D.}~\bibnamefont {Henderson}},
  \bibinfo {author} {\bibfnamefont {P.}~\bibnamefont {Plaschko}}, \ and\
  \bibinfo {author} {\bibfnamefont {W.}~\bibnamefont {Ronald~Fawcett}},\
  }\bibfield  {title} {\enquote {\bibinfo {title} {{Monte Carlo} and density
  functional theory study of the electrical double layer: the dependence of the
  charge/voltage relation on the diameter of the ions},}\ }\href@noop {}
  {\bibfield  {journal} {\bibinfo  {journal} {Mol. Simul.}\ }\textbf {\bibinfo
  {volume} {30}},\ \bibinfo {pages} {137--141} (\bibinfo {year}
  {2004})}\BibitemShut {NoStop}%
\bibitem [{\citenamefont {Gonz{\'a}lez-Tovar}\ \emph
  {et~al.}(2004)\citenamefont {Gonz{\'a}lez-Tovar}, \citenamefont
  {Jim{\'e}nez-{\'A}ngeles}, \citenamefont {Messina},\ and\ \citenamefont
  {Lozada-Cassou}}]{gonzalez04}%
  \BibitemOpen
  \bibfield  {author} {\bibinfo {author} {\bibfnamefont {E.}~\bibnamefont
  {Gonz{\'a}lez-Tovar}}, \bibinfo {author} {\bibfnamefont {F.}~\bibnamefont
  {Jim{\'e}nez-{\'A}ngeles}}, \bibinfo {author} {\bibfnamefont
  {R.}~\bibnamefont {Messina}}, \ and\ \bibinfo {author} {\bibfnamefont
  {M.}~\bibnamefont {Lozada-Cassou}},\ }\bibfield  {title} {\enquote {\bibinfo
  {title} {A new correlation effect in the {Helmholtz} and surface potentials
  of the electrical double layer},}\ }\href@noop {} {\bibfield  {journal}
  {\bibinfo  {journal} {J. Chem. Phys.}\ }\textbf {\bibinfo {volume} {120}},\
  \bibinfo {pages} {9782--9792} (\bibinfo {year} {2004})}\BibitemShut {NoStop}%
\bibitem [{\citenamefont {Guerrero-Garc{\'\i}a}\ \emph
  {et~al.}(2005)\citenamefont {Guerrero-Garc{\'\i}a}, \citenamefont
  {Gonz{\'a}lez-Tovar}, \citenamefont {Lozada-Cassou},\ and\ \citenamefont
  {de~J.~Guevara-Rodr{\'\i}guez}}]{guerrero05}%
  \BibitemOpen
  \bibfield  {author} {\bibinfo {author} {\bibfnamefont {G.~I.}\ \bibnamefont
  {Guerrero-Garc{\'\i}a}}, \bibinfo {author} {\bibfnamefont {E.}~\bibnamefont
  {Gonz{\'a}lez-Tovar}}, \bibinfo {author} {\bibfnamefont {M.}~\bibnamefont
  {Lozada-Cassou}}, \ and\ \bibinfo {author} {\bibfnamefont {F.}~\bibnamefont
  {de~J.~Guevara-Rodr{\'\i}guez}},\ }\bibfield  {title} {\enquote {\bibinfo
  {title} {The electrical double layer for a fully asymmetric electrolyte
  around a spherical colloid: An integral equation study},}\ }\href@noop {}
  {\bibfield  {journal} {\bibinfo  {journal} {J. Chem. Phys.}\ }\textbf
  {\bibinfo {volume} {123}},\ \bibinfo {pages} {034703} (\bibinfo {year}
  {2005})}\BibitemShut {NoStop}%
\bibitem [{\citenamefont {Partenskii}\ and\ \citenamefont
  {Jordan}(2005)}]{partenskii05}%
  \BibitemOpen
  \bibfield  {author} {\bibinfo {author} {\bibfnamefont {M.}~\bibnamefont
  {Partenskii}}\ and\ \bibinfo {author} {\bibfnamefont {P.}~\bibnamefont
  {Jordan}},\ }\bibfield  {title} {\enquote {\bibinfo {title} {Negative
  capacitance and instability at electrified interfaces: Lessons from the study
  of membrane capacitors},}\ }\href@noop {} {\bibfield  {journal} {\bibinfo
  {journal} {Condens. Matter Phys.}\ }\textbf {\bibinfo {volume} {8}},\
  \bibinfo {pages} {397--412} (\bibinfo {year} {2005})}\BibitemShut {NoStop}%
\end{thebibliography}

%
\end{document}